\documentclass[english,aps,notitlepage,nofootinbib,tightenlines,11pt,titlepage]{revtex4-1}
\pdfoutput=1
\usepackage[T1]{fontenc}
\usepackage[latin9]{inputenc}
\usepackage{array}
\usepackage{multirow}
\usepackage{amsmath}
\usepackage{amssymb}
\usepackage{graphicx}
\usepackage{wasysym}
\usepackage{esint}

\makeatletter

\providecommand{\tabularnewline}{\\}

\pdfoutput=1
\usepackage{afterpage}
\usepackage[colorlinks=true,citecolor=blue,linkcolor=black]{hyperref}
\usepackage{slashed}
\usepackage[normalem]{ulem}

\allowdisplaybreaks

\makeatother

\usepackage{babel}
\begin{document}

\title{Dirac neutrinos and $N_{{\rm eff}}$ II: the freeze-in case}

\author{Xuheng Luo$^{a}$, Werner Rodejohann$^b$ and Xun-Jie Xu$^{b,c}$}

\affiliation{$^{a}$Department of Physics and Astronomy, Johns Hopkins University
3400 North Charles Street, Baltimore, MD 21218, United States  \\
$^{b}$Max-Planck-Institut f\"ur Kernphysik, Postfach 103980, D-69029
Heidelberg, Germany\\
$^{c}$Service de Physique Th\'{e}orique, Universit\'{e} Libre de Bruxelles, Boulevard du Triomphe, CP225, 1050 Brussels, Belgium
}

\date{\today}
\begin{abstract}
\noindent
We discuss Dirac neutrinos whose right-handed component $\nu_R$ has new interactions that may lead to a measurable contribution to the effective number of relativistic neutrino species $N_{\rm eff}$. We aim at a model-independent and comprehensive study on a variety of possibilities. Processes for $\nu_R$-genesis from decay or scattering of thermal species, with spin-0, spin-1/2, or spin-1 initial or final states are all covered. We calculate numerically and analytically the contribution of $\nu_R$ to $N_{\rm eff}$ primarily in the freeze-in regime,  since the freeze-out regime has been studied before. While our approximate analytical results apply only to freeze-in,  our numerical calculations work for freeze-out as well, including the transition between the two regimes.
Using current and future constraints on $N_{\rm eff}$, we obtain limits and sensitivities of CMB experiments on masses and couplings of the new interactions. As a by-product, we obtain the contribution of Higgs-neutrino interactions,
$\Delta N_{\rm eff}^{\rm SM} \approx 7.5\times10^{-12}$, assuming the neutrino mass is 0.1 eV and generated by the standard Higgs mechanism.

\end{abstract}
\maketitle

\tableofcontents

\newpage
\section{Introduction}
\noindent 
While the knowledge of the neutrino parameters has increased in recent years, the two most important aspects have not been pinned down yet. That is, the absolute mass scale and the question whether light neutrinos are self-conjugate or not. The neutrino mass scale is only bounded from above \cite{Aker:2019uuj}, and both the Dirac and the Majorana character of neutrinos are compatible with all observations \cite{Dolinski:2019nrj}. 
Here we will assume that they are not self-conjugate, hence neutrinos are Dirac particles. 
The necessary presence of the right-handed components $\nu_R$ in this case  introduces the possibility that they contribute to the effective number of relativistic neutrino species $N_{\rm eff}$~\cite{Steigman:1979xp,Olive:1980wz,Dolgov:2002wy}. While in the Standard Model (SM) the contribution via Higgs-neutrino interactions is tiny (as we will confirm as a by-product of our study), new interactions of Dirac neutrinos can easily increase it to measurable sizes. This exciting possibility has been considered in several recent studies~\cite{Borah:2018gjk,Abazajian:2019oqj,Jana:2019mez,Calle:2019mxn,Luo:2020sho,Borah:2020boy,Adshead:2020ekg}\footnote{
In addition to this possibility,  a variety of other neutrino-related new physics could also affect $N_{\rm eff}$---see, e.g., ~\cite{Boehm:2012gr,Kamada:2015era,deSalas:2016ztq,Kamada:2018zxi,Escudero:2018mvt,Depta:2019lbe,Lunardini:2019zob,Escudero:2020dfa}. 
}.



In general, the contribution of $\nu_R$ to $N_{\rm eff}$  depends on both the coupling strength and the energy scale of the new interactions. If the energy scale is high and the coupling strength sizable,  $\nu_R$ are in thermal equilibrium with the dense and hot SM plasma at high temperatures. As the Universe cools down, the interaction rate decreases substantially due to the low densities and temperatures of  $\nu_R$ and the SM particle species. When the interaction rate can no longer keep up with the Universe's expansion, $\nu_R$ decouple from the SM plasma at a decoupling temperature $T_{\rm dec}$. Below $T_{\rm dec}$, the comoving entropy density of $\nu_R$ remains  a constant (i.e., $\nu_R$ freeze out), 
which fixes the contribution of $\nu_R$ to $N_{\rm eff}$. If all three flavors of $\nu_R$ decouple at a temperature much higher than the electroweak scale, their contribution to $N_{\rm eff}$ is 0.14 \cite{Dolgov:2002wy,Abazajian:2019oqj}, which is close to present constraints \cite{Akrami:2018vks,Aghanim:2018eyx} and can easily be probed/excluded by upcoming surveys \cite{Benson:2014qhw,Abitbol:2019nhf,Abazajian:2016yjj,Abazajian:2019eic}. 

In Ref.\ \cite{Luo:2020sho} we have 
considered the most general effective four-fermion contact interactions of Dirac neutrinos with the SM fermions 
and their effect on $N_{\rm eff}$.  Those contact interactions are assumed to be valid 
above the decoupling temperature, which usually holds for 
heavy particles with sizeable couplings (e.g., TeV particles with $> {\cal O} (10^{-2})$ couplings). However, small masses and/or tiny couplings are also rather common in many models, making these assumptions invalid.  


In fact, if the interactions are mediated by very weakly coupled particles (like the SM Higgs-neutrino coupling),  the right-handed neutrinos may never be in thermal equilibrium with the SM plasma. Nevertheless, via feeble interaction 
slowly some contribution of $\nu_R$ to the energy density and hence $N_{\rm eff}$ is built up, before the production stops (or becomes ineffective) because of dilution of the ingredients for $\nu_R$-genesis. In particular, if $\nu_R$ are produced from massive particles,  the production rate becomes exponentially suppressed when the temperature is below their masses. Hence, the comoving entropy density of $\nu_R$ will also be frozen at a certain level.  This freeze-in mechanism, first discussed in the context of dark matter~\cite{Hall:2009bx}, is the content of the present paper. 

We will assume here the presence of new interactions of $\nu_R$ with some generic boson ($B$) and fermion ($F$) which may or may not be SM particles.
In the most general set-up, one of, or both, $B$ and $F$ may be in equilibrium. In all cases, the mass hierarchy of $B$ and $F$ defines the dominating process that generates the $\nu_R$ density and thus the contribution to $N_{\rm eff}$. All possible cases are considered in this work, except the case when both $B$ and $F$ are not in equilibrium. In this case, additional interactions of those particles would be required to generate the $\nu_R$ density, which is beyond the model-independent study envisaged here. 
The case of a massless fermion $F$  includes  $F$ being the left-handed component of the Dirac neutrino (which is in equilibrium due to its SM interactions), and is also automatically part of this analysis. 
We show in this paper that if decay (scattering) of new particles is the dominating freeze-in process, limits on the new coupling constants of order 
$10^{-9}$ ($10^{-4}$) may be constrained for new particle masses around GeV. Our framework also allows us to calculate the contribution of SM Dirac neutrinos to $N_{\rm eff}$, for which the freeze-in occurs via the tiny Yukawa interactions with the Higgs boson: $\Delta N_{\rm eff}^{\rm SM} \approx 7.5\times10^{-12} \, (m_{\nu}/(0.1\, {\rm eV}))^{2}$.\\ 

The paper is built up as follows: In Section \ref{sec:basic} we discuss our framework and the several cases that may be present. The calculation of the interaction rates is summarized in Section \ref{sec:Squared-amplitudes}. An analytical estimate of the resulting contribution to $N_{\rm eff}$ is given in Section \ref{sec:Analytic}, and compared to the numerical result for Dirac neutrino masses generated by the SM Higgs mechanism in Section \ref{sec:Higgs}.  
The full numerical analysis for the general cases is presented in Section \ref{sec:Numerical-results}. 
We conclude in Section \ref{sec:Conclusion} and put several technical details in Appendices.

\section{Framework\label{sec:basic}}

\noindent If neutrinos are Dirac particles and have beyond the Standard Model (BSM)  interactions,
generically one can consider the following Lagrangian\footnote{Throughout this paper, we assume that the new interactions of neutrinos universally couple to all flavors with flavor-independent coupling constants.}:
\begin{equation}
{\cal L}\supset g_{\nu}B\overline{F}\nu_{R}+{\rm h.c.},\label{eq:m-19}
\end{equation}
where $g_{\nu}$ is a coupling constant, $B$ and $F$ stand for a
scalar boson and a chiral fermion, respectively. Besides this scalar interaction, we also consider
the vector case:
\begin{equation}
{\cal L}\supset g_{\nu}B^{\mu}\overline{F}\gamma_{\mu}\nu_{R}+{\rm h.c.},\label{eq:m-20}
\end{equation}
for which the analysis will be  similar. In both
cases, the masses of $B$ and $F$ are denoted by $m_{B}$ and $m_{F}$, 
respectively.   Note that in our framework $B$ and $F$ can be
BSM or SM particles\footnote{In fact, if both $B$ and $F$ are SM particles, the only possible
interaction that can arise from a gauge invariant terms is $h\overline{\nu_{L}}\nu_{R}$
where $h$ is the SM Higgs (see Sec.~V). If one of them is a non-SM
particle, then it allows for more possibilities. Here we refrain from
further discussions on model-dependent details and concentrate on
the generic  framework.}.
What is essentially relevant here is whether
they are in thermal equilibrium or not during the $\nu_{R}$-genesis
epoch. 
Therefore we have the following  cases (see Tab.\ \ref{tab:Dominant-processes}):
\begin{itemize}
\item (I) Both $B$ and $F$ are  in thermal equilibrium. In this case,
the dominant process for $\nu_{R}$-genesis is $B$ or $F$ decay:
$B\rightarrow F+\overline{\nu_{R}}$  (if $m_{B}>m_{F}$) or $F\rightarrow B+\nu_{R}$
(if $m_{F}>m_{B}$), to which we refer as subcases (I-1) and (I-2)
respectively.
Note that other processes such as $B+\overline{B}\rightarrow\nu_{R}+\overline{\nu_{R}}$
and $F+\overline{F}\rightarrow\nu_{R}+\overline{\nu_{R}}$ also contribute
to $\nu_{R}$-genesis. Being typically a factor of $g_{\nu}^{2}/(16\pi^{2})$
smaller than the decay processes, their contributions in this case
are subdominant.
\item (II) Only $B$ is in thermal equilibrium while $F$ is not. If $B$
is heavier than $F$, defined as subcase (II-1), then the dominant
process for $\nu_{R}$-genesis is still $B$ decay, similar to (I-1).
We should note, however, that the collision term in (II-1) is different
from that of (I-1), as will be shown later  in Eqs.~(\ref{eq:m-92})-(\ref{eq:m-97}). 
If $F$ is heavier than
$B$, since $F$ is assumed not to be in thermal equilibrium, $F$ decay  is  less
productive than  $B$ annihilation:  $B+\overline{B}\rightarrow\nu_{R}+\overline{\nu_{R}}$
via the $t$-channel diagram in Tab.~\ref{tab:Dominant-processes}.
We refer to it as subcase (II-2).
\item (III) Only $F$ is in thermal equilibrium while $B$ is not. Likewise,
we have subcase (III-1) for $m_{F}>m_{B}$ and subcase (III-2) for
$m_{B}>m_{F}$, with their dominant processes being $F\rightarrow B+\nu_{R}$
and $F+\overline{F}\rightarrow\nu_{R}+\overline{\nu_{R}}$,  respectively. 
\item (IV) Neither $F$ or $B$ is in thermal equilibrium. If in a Dirac neutrino
model, given a new interaction in Eq.~(\ref{eq:m-19}) or (\ref{eq:m-20}),
neither of them is in thermal equilibrium, one should check
whether there are other interactions involving different 
particles, 
which would be the dominant contribution
to $\nu_{R}$ production. If indeed all interactions of $\nu_{R}$
in the model are in case (IV), then  typically the abundance of $\nu_{R}$
is suppressed.
Although if neither of them is in thermal equilibrium, sizable abundances of $F$, $B$ and hence $\nu_{R}$
are still possible, quantitative results in this case depend however not
only on $g_{\nu}$ but also on other parameters (e.g.\ the couplings
of $F$ and $B$ to the SM content). Hence we leave this model-dependent
case to future work.
\end{itemize}
We summarize the above cases in Tab.~\ref{tab:Dominant-processes}. Note that we will remain agnostic about the origin of the above two interactions in Eqs.\ (\ref{eq:m-19}) and (\ref{eq:m-20}). Without a full-fledged UV-complete model there may arise conceptual issues for the vector case, which will be discussed later. In addition, if $B$ or $F$ are sufficiently light, they may also contribute to $N_{\rm eff}$ directly (see, e.g.,~\cite{Huang:2017egl,Berbig:2020wve,He:2020zns}), depending on whether they are SM particles or not, and on their thermal evolution. This possibility will not be studied in this work.
\afterpage{
\clearpage 

\begin{table*}[h!]
\caption{Dominant processes for $\nu_{R}$-genesis in the $B\overline{F}\nu_{R}$
framework---see Eqs.~(\ref{eq:m-19}) and (\ref{eq:m-20}) and discussions
below. For the vector case, dashed lines are interpreted as vector
bosons. Some expressions use the Mandelstam parameters $s$, $t$,
$u$. 
To
avoid IR divergences, some results are only valid for $16\pi^{2}m_{B}^{2} \gtrsim g_\nu^2 m_{F}^{2} $
(see text for more details).
 \label{tab:Dominant-processes} 
 }

\begin{ruledtabular}
\begin{tabular}{ccc}
Cases & Dominant processes for $\nu_{R}$-genesis & $S|{\cal M}|^{2}$\tabularnewline
\hline 
 & \multirow{6}{*}{\includegraphics[width=4cm]{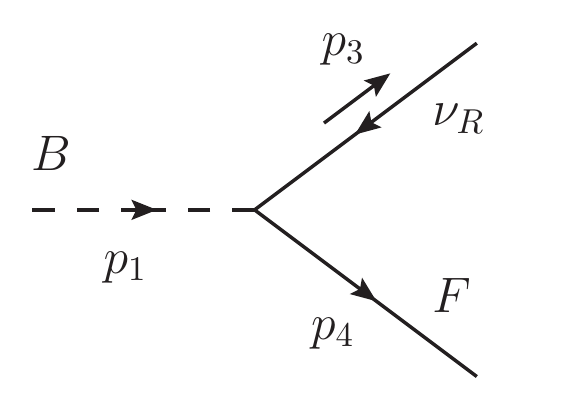}} & scalar $B$:\tabularnewline
 &  &   $|g_{\nu}|^{2}(m_{B}^{2}-m_{F}^{2})$\tabularnewline
(I-1) $F$ and $B$ in &  & \tabularnewline
thermal equilibrium, &  & vector $B^{\mu}$:\tabularnewline
$m_{B}>m_{F}$ &  & $|g_{\nu}|^{2}\left(2m_{B}^{2}-m_{F}^{2}-\frac{m_{F}^{4}}{m_{B}^{2}}\right)$\tabularnewline
 &  & \tabularnewline
\hline 
 & \multirow{6}{*}{\includegraphics[width=4cm]{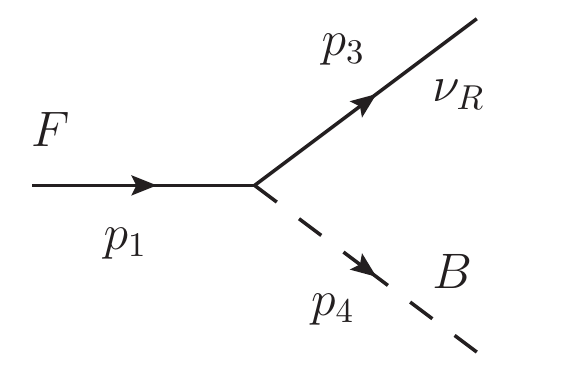}} & scalar $B$:\tabularnewline
 &  &  $|g_{\nu}|^{2}(m_{F}^{2}-m_{B}^{2})$\tabularnewline
(I-2) $F$ and $B$ in &  & \tabularnewline
thermal equilibrium, &  & vector $B^{\mu}$ (for $16\pi^{2}m_{B}^{2}\gtrsim g_\nu^2 m_{F}^{2}$):\tabularnewline
$m_{F}>m_{B}$ &  &  $|g_{\nu}|^{2}(m_{F}^{2}-m_{B}^{2})(2m_{B}^{2}+m_{F}^{2})m_{B}^{-2}$\tabularnewline
 &  & \tabularnewline
\hline 
 & \multirow{6}{*}{\includegraphics[width=4cm]{fig/B_decay}} & scalar $B$:\tabularnewline
 &  &  $|g_{\nu}|^{2}(m_{B}^{2}-m_{F}^{2})$\tabularnewline
(II-1) $B$ in thermal &  & \tabularnewline
equilibrium, $F$ not, &  & vector $B^{\mu}$:\tabularnewline
$m_{B}>m_{F}$ &  & $\frac{|g_{\nu}|^{2}}{3m_{B}^{2}}(m_{B}^{2}-m_{F}^{2})(2m_{B}^{2}+m_{F}^{2})$\tabularnewline
 &  & \tabularnewline
\hline 
 & \multirow{6}{*}{\includegraphics[width=4cm]{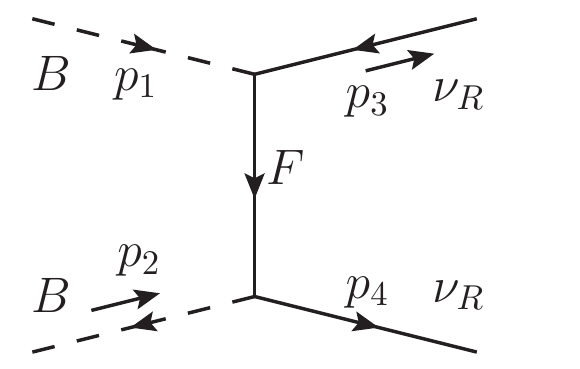}} & complex scalar $B$:\tabularnewline
 &  & $|g_{\nu}|^{4}\frac{tu-m_{B}^{4}}{|t-m_{F}^{2}|^{2}}$\tabularnewline
(II-2) $B$ in thermal &  & \tabularnewline
equilibrium, $F$ not, &  & for real scalar $B$, see Eq.~(\ref{eq:m-82})\tabularnewline
$m_{F}>m_{B}$ &  & for vector $B$, see Eqs.~(\ref{eq:m-41}) and (\ref{eq:m-84})\tabularnewline
 &  & \tabularnewline
\hline 
 & \multirow{6}{*}{\includegraphics[width=4cm]{fig/F_decay}} & scalar $B$:\tabularnewline
 &  &  $|g_{\nu}|^{2}(m_{F}^{2}-m_{B}^{2})$\tabularnewline
(III-1) $F$ in thermal &  & \tabularnewline
equilibrium, $B$ not, &  & vector $B^{\mu}$ (for $16\pi^{2}m_{B}^{2}\gtrsim g_\nu^2 m_{F}^{2}$):\tabularnewline
$m_{F}>m_{B}$ &  &  $|g_{\nu}|^{2}(m_{F}^{2}-m_{B}^{2})(2m_{B}^{2}+m_{F}^{2})m_{B}^{-2}$\tabularnewline
 &  & \tabularnewline
\hline 
 & \multirow{6}{*}{\includegraphics[width=4cm]{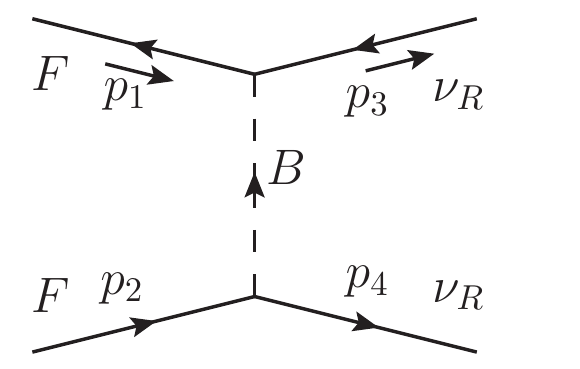}} & scalar $B$:\tabularnewline
 &  & $|{\cal M}|^{2}=|g_{\nu}|^{4}(t-m_{F}^{2})^{2}/(t-m_{B}^{2})^{2}$\tabularnewline
(III-2) $F$ in thermal &  & \tabularnewline
equilibrium, $B$ not, &  & vector $B^{\mu}$:\tabularnewline
$m_{B}>m_{F}$ &  &  $4|g_{\nu}|^{4}(m_{F}^{2}-u)^{2}/(t-m_{B}^{2})^{2}$\tabularnewline
 &  & \tabularnewline
\hline 
 &  & \tabularnewline
(IV) $B$ \& $F$ not in & Model-dependent; & \tabularnewline
thermal equilibrium & Abundance of $\nu_{R}$ usually  suppressed & \tabularnewline
 &  & \tabularnewline
\end{tabular}\end{ruledtabular}

\end{table*}

\clearpage
}

\mbox{ }\\
The $\nu_{R}$ energy density, $\rho_{\nu_{R}}$, is  determined by
the following Boltzmann equation \cite{Luo:2020sho}:
\begin{equation}
\dot{\rho}_{\nu_{R}}+4H\rho_{\nu_{R}}=C_{\nu_{R}}.\label{eq:m-22}
\end{equation}
Here $\dot{\rho}_{\nu_{R}}\equiv d\rho_{\nu_{R}}/dt$, $H$ is the
Hubble parameter, and $C_{\nu_{R}}$ is  referred to as the collision
term. For a $2\rightarrow2$ process, the collision term is computed
from the following integral:
\begin{eqnarray}
C_{\nu_{R}} & \equiv & N_{\nu_{R}}\int E_{\nu_{R}}d\Pi_{1}d\Pi_{2}d\Pi_{3}d\Pi_{4}(2\pi)^{4}\delta^{4}(p_{1}+p_{2}-p_{3}-p_{4})\nonumber \\
 &  & \times S|{\cal M}|^{2}\left[f_{1}f_{2}(1\pm f_{3})(1\pm f_{4})-f_{3}f_{4}(1\pm f_{1})(1\pm f_{2})\right],\label{eq:m-21}
\end{eqnarray}
\begin{equation}
d\Pi_{i}\equiv\frac{1}{(2\pi)^{3}}\frac{d^{3}p_{i}}{2E_{i}},\ \ f_{i}\equiv\frac{1}{\exp\left(E_{i}/T_{i}\right)\mp1},\ \ (i=1,\thinspace2,\thinspace3,\thinspace4),\label{eq:m-81}
\end{equation}
 where $N_{\nu_{R}}=6$ (including  $\nu$ and $\overline{\nu}$ of
three flavors\footnote{Conceptually, we treat particles and anti-particles as different species
in the thermal plasma rather than the same species with doubled internal
degrees of freedom. This treatment can simplify  a few potential issues
related to the symmetry factor and conjugate processes (e.g., whether
$F\rightarrow B+\nu_{R}$ and $\overline{F}\rightarrow\overline{B}+\overline{\nu_{R}}$
should be taken into account simultaneously or not). In practice,
due to the identical thermal distributions, we combine them into a
single equation so that $\rho_{\nu_{R}}$ in Eq.~(\ref{eq:m-22})
contains the energy density of both $\nu_{R}$ and $\overline{\nu_{R}}$.
For more detailed discussions on this issue, see Ref.~\cite{Luo:2020sho}.}); $E_{\nu_{R}}$ is the energy of $\nu_{R}$; $S$ is the symmetry
factor (which in most cases\footnote{The only exception here is subcase (II-2) when $B$ is a real field.
More details will be discussed when $|{\cal M}|^{2}$ is computed.
} is $1$); $|{\cal M}|^{2}$ is the squared amplitude of the process;
$p_{i}$, $E_{i}$, and $T_{i}$ denote the momentum, energy, and
temperature of the $i$-th particle in the process. To be more specific,
we have labeled the momenta $p_{1}$, $p_{2}$, $p_{3}$ and $p_{4}$
for each $2\rightarrow2$ process in Tab.~\ref{tab:Dominant-processes}.  
For decay processes presented in Tab.~\ref{tab:Dominant-processes} 
we avoid using $p_{2}$, hence the final momenta are still $p_{3}$ and
$p_{4}$, as already indicated in the diagrams  in Tab.~\ref{tab:Dominant-processes}.
In this way, one can apply Eq.~(\ref{eq:m-21}) to decay processes
with a minimal modification: only quantities with subscripts ``$2$''
need to be removed. In addition, since in all the diagrams $p_{3}$
is always the momentum of $\nu_{R}$, we set $E_{\nu_{R}}=E_{3}$
in Eq.~(\ref{eq:m-21}).

In the presence of energy injection to the $\nu_{R}$ sector,   the
SM sector obeys the following Boltzmann equation:
\begin{equation}
\dot{\rho}_{{\rm SM}}+3H(\rho_{{\rm SM}}+P_{{\rm SM}})=-C_{\nu_{R}},\label{eq:m-23}
\end{equation}
where $\rho_{{\rm SM}}$ and $P_{{\rm SM}}$ are the energy density
and pressure of SM particles. In later discussions, we may also use
the entropy density of the SM, denoted by $s_{{\rm SM}}\equiv(\rho_{{\rm SM}}+P_{{\rm SM}})/T$.
The three thermal quantities have the following temperature dependence:
\begin{equation}
\rho_{{\rm SM}}=g_{\star}^{(\rho)}\frac{\pi^{2}}{30}T^{4},\ \ P_{{\rm SM}}=g_{\star}^{(P)}\frac{\pi^{2}}{90}T^{4},\ \ s_{{\rm SM}}=g_{\star}^{(s)}\frac{2\pi^{2}}{45}T^{3}.\label{eq:m-24}
\end{equation}
The effective degrees of freedom of the SM, namely $g_{\star}^{(\rho)}$,
$g_{\star}^{(P)}$, and $g_{\star}^{(s)}$, can reach $106.75$ at
sufficiently high temperatures, and for $T$ at a few MeV are almost
equal to  $10.75$, coming from three left-handed neutrinos, two chiral
electrons, and one photon: $2\times3\times7/8+2\times2\times7/8+2=43/4$.
We refer to  Fig.\ 2.2 in Ref.\ \cite{Baumann:2019nls} 
for recent calculations of
$g_{\star}^{(\rho)}$ 
which will be used in
our analyses. 
Regarding the small difference between $g_{\star}^{(P)}$ and $g_{\star}^{(\rho)}$ which is important for entropy conservation,
we use $d g_{\star}^{(P)}/dT=3(g_{\star}^{(\rho)}-g_{\star}^{(P)})/T$~\cite{Luo:2020sho} to obtain $g_{\star}^{(P)}$ from $g_{\star}^{(\rho)}$.

In this work, we  study the effect of Dirac neutrinos on $N_{{\rm eff}}$
by solving Eqs.~(\ref{eq:m-22}) and (\ref{eq:m-23}) analytically
(see Sec.~\ref{sec:Analytic}) or numerically (see Sec.~\ref{sec:Numerical-results}).
When the solution is obtained, the $\nu_{R}$ contribution  to $N_{{\rm eff}}$
can be computed by
\begin{equation}
\Delta N_{{\rm eff}}=\frac{4}{7}g_{\star,{\rm dec}}^{(\rho)}\left[\frac{10.75}{g_{\star,{\rm dec}}^{(s)}}\right]^{4/3}\frac{\rho_{\nu_{R},{\rm dec}}}{\rho_{{\rm SM},{\rm dec}}},\label{eq:m-87}
\end{equation}
where the subscript ``dec'' denotes any moment after $\nu_{R}$
is fully decoupled from the SM plasma.  In practical use, one only
needs to solve Eqs.~(\ref{eq:m-22}) and (\ref{eq:m-23}) starting
at a sufficiently high temperature and ending at any low temperature
that is much smaller than $m_{F}$ or $m_{B}$, because at such temperatures
$C_{\nu_{R}}$ no longer makes significant contributions. More practically,
because $g_{\star}^{(\rho)}\approx g_{\star}^{(s)}\approx10.75$ when
$T$ is about a few MeV, Eq.~(\ref{eq:m-87}) can be reduced to
\begin{equation}
\Delta N_{{\rm eff}}\approx N_{\nu}\left(\frac{T_{\nu_{R},{\rm low}}}{T_{{\rm low}}}\right)^{4},\label{eq:m-88}
\end{equation}
where $N_{\nu}=3$ and the subscript ``low'' denotes generally any moment at which the
approximation $g_{\star}^{(\rho)}\approx g_{\star}^{(s)}\approx10.75$
is valid, typically between 5 and 10 MeV (at $T=10$ MeV, $g_{\star}^{(\rho)}\approx g_{\star}^{(s)}\approx10.76$
and at $T=5$ MeV, $g_{\star}^{(\rho)}\approx g_{\star}^{(s)}\approx10.74$~\cite{Husdal:2016haj}).

\section{Squared amplitudes\label{sec:Squared-amplitudes}}

\noindent To proceed with the analyses on the various cases summarized
in Tab.~\ref{tab:Dominant-processes}, we need to compute the squared
amplitude $|{\cal M}|^{2}$ for each dominant process and take  the
symmetry factors into account properly. The result is summarized  in Tab.~\ref{tab:Dominant-processes}.

\subsection{$B$ decay (scalar case)}

\noindent This is the dominant process of $\nu_R$-genesis for subcases (I-1)
and (II-1), assuming $B$ is a scalar boson.
The squared amplitude of scalar $B$ decay  
reads:
\begin{equation}
|{\cal M}|^{2}=\sum_{s_{4},\thinspace s_{3}}
|g_{\nu}\overline{u_{4}}P_{R}v_{3}|^{2} = 
2|g_{\nu}|^{2}\left(p_{3}\cdot p_{4}\right) = 
|g_{\nu}|^{2}(m_{B}^{2}-m_{F}^{2}),\label{eq:m-34}
\end{equation}
where $v_{3}$ and $u_{4}$ denote the final fermionic states. 
In the second ``$=$'', we have applied the standard trace technology to the spin sum of $s_{3}$ and $s_{4}$ 
Note that due to the projector $P_{R}$ in Eq.~(\ref{eq:m-34}), only
right-handed neutrinos and left-handed $F$ are included. Despite
being formally included in the summation of $s_{3}$ and $s_{4}$,
contributions of left-handed neutrinos and right-handed $F$ automatically
vanish. 
In the third  ``$=$'', we have used on-shell conditions. 
More specifically (and also for later use in other cases), 
we can expand $p_{1}^{2}=(p_{3}+p_{4})^{2}$,
$p_{4}^{2}=(p_{1}-p_{3})^{2}$, and $p_{3}^{2}=(p_{1}-p_{4})^{2}$ to obtain
\begin{eqnarray}
p_{3}\cdot p_{4} & = & (m_{1}^{2}-m_{3}^{2}-m_{4}^{2})/2,\label{eq:m-56}\\
p_{1}\cdot p_{3} & = & (m_{1}^{2}+m_{3}^{2}-m_{4}^{2})/2,\label{eq:m-57}\\
p_{1}\cdot p_{4} & = & (m_{1}^{2}-m_{3}^{2}+m_{4}^{2})/2,\label{eq:m-58}
\end{eqnarray}
where $m_1$, $m_3$, and $m_4$ are the masses of particles 1, 3, and 4, respectively. 
For the current process, we have $m_1=m_B$, $m_3=0$, and $m_4=m_F$.

\subsection{$B$ decay (vector case)\label{sec:B}}

\noindent This is the dominant process of $\nu_R$-genesis for subcases (I-1)
and (II-1), assuming $B$ is a vector boson.
The squared amplitude is similar to the previous one, execpt that here we add a polarization vector
$\epsilon^{\mu}$ and a $\gamma_{\mu}$:
\begin{equation}
|{\cal M}|^{2}=\sum_{\epsilon}\sum_{s_{4},\thinspace s_{3}}|g_{\nu}\epsilon^{\mu}\overline{u_{4}}\gamma_{\mu}P_{R}v_{3}|^{2}.\label{eq:m-36}
\end{equation}
Since the vector boson is in initial states,  in principle, we would
need to take the average over vector polarizations, which would imply
that Eq.~(\ref{eq:m-36}) should be divided by a factor of three.
However, since a massive vector boson has three internal degrees
of freedom and each degree of freedom contributes equally to $C_{\nu_{R}}$,
we would have to multiply the integrand in Eq.~(\ref{eq:m-21}) by
a factor of three; or alternatively, the factor of three should be
included in $d\Pi$ in Eq.~(\ref{eq:m-81}). To keep Eqs.~(\ref{eq:m-21})
and (\ref{eq:m-81}) in their current form, we do not add the factor
of three in $|{\cal M}|^{2}$. As aforementioned, conceptually, we
treat each internal degree of a particle as an independent thermal
species. Hence $|{\cal M}|^{2}$ in Eq.~(\ref{eq:m-36}) should be
interpreted as the total squared amplitude of the three species decaying
to $\nu_{R}$ and $F$. 

 When summing over vector polarization, we need
\begin{equation}
\sum_{\epsilon}\epsilon_{\mu}(q)\epsilon_{\nu}^{*}(q)=\frac{q_{\mu}q_{\nu}}{m_{B}^{2}}-g_{\mu\nu}.\label{eq:m-33}
\end{equation}
Hence, after performing the summation of spins and  vector polarization, we obtain
\begin{eqnarray}
|{\cal M}|^{2} & = & |g_{\nu}|^{2}\sum_{\epsilon}\epsilon_{\mu}\epsilon_{\nu}^{*}\, {\rm tr\!}\left[(\slashed{p}_{4}+m_{4})\gamma^{\mu}P_{R}\slashed{p}_{3}P_{L}\gamma_{\nu}\right]\nonumber \\
 & = & |g_{\nu}|^{2}\left(2m_{B}^{2}-m_{F}^{2}-\frac{m_{F}^{4}}{m_{B}^{2}}\right),
 \label{eq:m-37}
\end{eqnarray}
where we have replaced scalar products of $p_1$ with $p_3$ and $p_4$ with particle masses 
according to Eqs.~(\ref{eq:m-56})-(\ref{eq:m-58}).

\subsection{$F$ decay (scalar case)}

\noindent This is the dominant process of $\nu_R$-genesis for subcases (I-2)
and (III-1), assuming $B$ is a scalar boson.
For these two subcases, the diagram shown in Tab.~\ref{tab:Dominant-processes}
is generated by $g_{\nu}^{*}B^{\dagger}\overline{\nu_{R}}F=g_{\nu}^{*}B^{\dagger}\overline{\nu_{R}}P_{L}F$
instead of $g_{\nu}B\overline{F}\nu_{R}$. Hence the squared amplitude
reads:
\begin{equation}
|{\cal M}|^{2}=\sum_{s_{1},\thinspace s_{3}}|g_{\nu}^{*}\overline{u_{3}}P_{L}u_{1}|^{2}
=2|g_{\nu}|^{2}\left(p_{1}\cdot p_{3}\right)
=|g_{\nu}|^{2}(m_{F}^{2}-m_{B}^{2})
,\label{eq:m-34-1}
\end{equation}
where $u_{1}$ is the initial fermionic state. Note that due to the
chiral projector $P_{L}$, only left-handed $F$ can decay to $\nu_{R}$.
Therefore, the process can be treated either as unpolarized $F$ decay,
which would contain a factor of $1/2$ in Eq.~(\ref{eq:m-34-1}),
or as polarized $F$ decay (left-handed), which does not contain such
a factor. Although conceptually different, the two approaches are
equivalent. When computing the collision term, the factor of $1/2$
in the unpolarized approach would be canceled  by an additional
factor of 2 in the integrand due to the inclusion of the right-handed
component of  $F$. Here we adopt the polarized approach because
in some models where $F$ is a chiral fermion its right-handed component
is absent.


\subsection{$F$ decay (vector case)\label{sec:D}}

\noindent This is the dominant process of $\nu_R$-genesis for subcases (I-2)
and (III-1), assuming $B$ is a vector boson.
Similar to the previous calculation, 
we add a
polarization vector $\epsilon^{\mu}$ in Eq.~(\ref{eq:m-34-1}) and
sum over it according to  Eq.~(\ref{eq:m-33}). Therefore, the squared amplitude reads
\begin{eqnarray}
|{\cal M}|^{2} & = & \sum_{\epsilon}\sum_{s_{4},\thinspace s_{3}}|g_{\nu}^{*}\epsilon_{\mu}^{*}\overline{u_{3}}P_{L}\gamma^{\mu}u_{1}|^{2}\nonumber \\
 & = & |g_{\nu}|^{2}\sum_{\epsilon}\epsilon_{\mu}\epsilon_{\nu}^{*}\, {\rm tr\! }\left[\slashed{p}_{3}P_{L}\gamma^{\mu}(\slashed{p}_{1}+m_{1})\gamma_{\nu}P_{R}\right]\nonumber \\
 & = & 2|g_{\nu}|^{2}\left[p_{1}\cdot p_{3}+\frac{2(p_{1}\cdot p_{4})(p_{3}\cdot p_{4})}{m_{B}^{2}}\right] \nonumber \\
 & = & |g_{\nu}|^{2}\frac{(m_{F}^{2}-m_{B}^{2})(2m_{B}^{2}+m_{F}^{2})}{m_{B}^{2}}.\label{eq:m-63}
\end{eqnarray}
Here we would like to discuss the IR divergence $m_{B}\rightarrow0$
in the above result. The divergence of $m_{B}\rightarrow0$ was already present 
in Eq.~(\ref{eq:m-33}).
Recall that in unbroken $U(1)$ gauge
theories we have the Ward identity $q^{\mu}{\cal M}_{\mu}=0$ for
any Feynman diagram with a photon external leg ($\epsilon^{\mu}$)
being replaced by $q^{\mu}$. Therefore, whenever the Ward identity
is valid, the longitudinal part $q^{\mu}q^{\nu}$ in 
Eq.~(\ref{eq:m-33})
has no contribution. In our framework, we consider a generic interaction 
($B^{\mu}F\gamma_{\mu}\nu_{R}$) without specifying the origin of
the gauge boson mass. In this case, the Ward identity is in general
not valid and the cancellation of the IR divergence becomes quite
model dependent. In fact, when $m_{B}$ is small, generally one should
not expect a strong hierarchy between $m_{F}$ and $m_{B}$ because
the self-energy diagram of $B^{\mu}$ generated by two $g_{\nu}B^{\mu}F\gamma_{\mu}\nu_{R}$
vertices is of ${\cal O}(g_{\nu}^{2}m_{F}^{2}/16\pi^{2})$. Thus, a strong
mass hierarchy such as $m_{B}^{2}/m_{F}^{2}\ll g_{\nu}^{2}/16\pi^{2}$ would
be unstable under loop corrections. As a rule of thumb, we suggest
that Eq.~(\ref{eq:m-63}) should be used only when $m_{B}$ is in
the regime of  $m_{F}^{2}>m_{B}^{2}\apprge g_{\nu}^{2} m_{F}^{2}/16\pi^{2}$. 

\subsection{$B$ annihilation (scalar case)}

\noindent 
This is the dominant process of $\nu_R$-genesis for subcase (II-2), assuming $B$ is a scalar boson.
Let us first consider complex $B$ so that the two
initial states are not identical particles. For complex $B$,  the
upper vertex of the Feynman diagram for subcase (II-2) is generated by $g_{\nu}B\overline{F}P_{R}\nu_{R}$,
and the lower vertex by its conjugate ($g_{\nu}^{*}B^{\dagger}\overline{\nu_{R}}P_{L}F$).
The squared amplitude reads:
\begin{eqnarray}
|{\cal M}|^{2} & = & \sum_{s_{4},\thinspace s_{3}}\left|g_{\nu}^{*}\overline{u_{4}}P_{L}\frac{i}{\slashed{p}_{F}-m_{F}}g_{\nu}P_{R}v_{3}\right|^{2}\label{eq:m-40} \\
& = & \frac{|g_{\nu}|^{4}}{|p_{F}^{2}-m_{F}^{2}|^{2}}\, {\rm tr\! }\left[\slashed{p}_{4}P_{L}(\slashed{p}_{F}+m_{F})P_{R}\slashed{p}_{3}P_{L}(\slashed{p}_{F}+m_{F})P_{R}\right]\nonumber \\
 & = & 2|g_{\nu}|^{4}\frac{2(p_{1}\cdot p_{3})(p_{1}\cdot p_{4})-m_{B}^{2}(p_{3}\cdot p_{4})}{|{p}_{F}^2-m_{F}^{2}|^{2}}\nonumber \\
 & = & |g_{\nu}|^{4}\frac{tu-m_{B}^{4}}{|t-m_{F}^{2}|^{2}}\thinspace,\label{eq:m-39}
\end{eqnarray}
where $p_{F}=p_{1}-p_{3}$ and we have used the usual Mandelstam parameters\footnote{We note that $t$ in this paper has been used to denote time as well
as a Mandelstam parameter (both are very standard notations). Potential
confusion can be avoided if we notice that the former has the dimension
of ${\rm [energy]}^{-1}$ and the latter has ${\rm [energy]}^{2}$.}: 
\begin{eqnarray}
s & \equiv & (p_{1}+p_{2})^{2}=(p_{3}+p_{4})^{2},\label{eq:m-42}\\
t & \equiv & (p_{1}-p_{3})^{2}=(p_{4}-p_{2})^{2},\label{eq:m-43}\\
u & \equiv & (p_{1}-p_{4})^{2}=(p_{3}-p_{2})^{2}.\label{eq:m-44}
\end{eqnarray}
In additon, we have used $s+t+u=\sum_{i}m_{i}^{2}$ to simplify the result in Eq.~(\ref{eq:m-39}). 



Next, we consider that $B$ is a real scalar which implies that the two initial
states can be interchanged. In this case, we actually have two diagrams.
The second diagram is obtained by interchanging the $p_{1}$ and $p_{2}$
lines. Due to identical particles, we have the symmetry factor $S=\frac{1}{2!}$. 
Therefore, Eq.~(\ref{eq:m-40})
should be modified as
\begin{equation}
S|{\cal M}|^{2}=\frac{1}{2!}\sum_{s_{4},\thinspace s_{3}}\left|g_{\nu}^{*}\overline{u_{4}}P_{L}\left[\frac{i}{\slashed{p}_{F}-m_{F}}+\frac{i}{\slashed{p}'_{F}-m_{F}}\right]g_{\nu}P_{R}v_{3}\right|^{2},\label{eq:m-40-2}
\end{equation}
where $p_{F}'=p_{2}-p_{3}$ is the momentum of $F$ in the second
diagram. 
Following a similar calculation, we obtain
\begin{equation}
S|{\cal M}|^{2}=\frac{|g_{\nu}|^{4}}{2}\left[\frac{(t-u)^{2}(tu-m_{B}^{4})}{\left(t-m_{F}^{2}\right)^{2}\left(u-m_{F}^{2}\right)^{2}}\right].\label{eq:m-82}
\end{equation}
As is expected, the full result is $p_{1}\leftrightarrow p_{2}$ (corresponding
to $t\leftrightarrow u$) symmetric because the two initial particles
are identical.

\subsection{$B$ annihilation (vector case)}

\noindent 
This is the dominant process of $\nu_R$-genesis for subcase (II-2), assuming $B$ is a vector boson.
As a vector field, for $B^{\mu}$ it is also possible to be complex
(similar to $W^{\pm}$  in the SM). For real $B^{\mu}$, again,
we need to be careful about the issue of identical particles. Let
us first consider complex $B$. In this case,  the upper and lower
vertices are generated by $g_{\nu}B^{\mu}\overline{F}\gamma_{\mu}P_{R}\nu_{R}$
and $g_{\nu}^{*}B^{*\mu}\overline{\nu_{R}}P_{L}\gamma_{\mu}F$. The
initial states contain two polarization vectors, denoted as $\epsilon_{1}^{\mu}$
and $\epsilon_{2}^{\mu}$. Hence we modify Eq.~(\ref{eq:m-40}) to
the following form:
\begin{equation}
|{\cal M}|^{2}=\sum_{\epsilon_{1},\thinspace\epsilon_{2}}\sum_{s_{4},\thinspace s_{3}}|g_{\nu}^{*}\epsilon_{2}^{\mu}\overline{u_{4}}P_{L}\gamma_{\mu}\frac{i}{\slashed{p}_{F}-m_{F}}g_{\nu}\epsilon_{1}^{\rho}\gamma_{\rho}P_{R}v_{3}|^{2},\label{eq:m-36-1}
\end{equation}
which gives
\begin{eqnarray}
|{\cal M}|^{2} & = & \frac{|g_{\nu}|^{4}}{|p_{F}^{2}-m_{F}^{2}|^{2}}\left[\sum_{\epsilon_{2}}\epsilon_{2}^{\mu}\epsilon_{2}^{*\nu}\right]\left[\sum_{\epsilon_{1}}\epsilon_{1}^{\rho}\epsilon_{1}^{*\sigma}\right]\nonumber \\
&  & \times\, {\rm tr\! }\left[\slashed{p}_{4}P_{L}\gamma_{\mu}(\slashed{p}_{F}+m_{F})\gamma_{\rho}P_{R}\slashed{p}_{3}P_{L}\gamma_{\sigma}(\slashed{p}_{F}+m_{F})\gamma_{\nu}P_{R}\right] \nonumber \\
 &= & \frac{|g_{\nu}|^{4}}{|t-m_{F}^{2}|^{2}}\left[\frac{t^{3}u}{m_{B}^{4}}-\frac{4t^{2}(t+u)}{m_{B}^{2}}-4m_{B}^{4}+t(7t+4u)\right].
 \label{eq:m-41}
\end{eqnarray}

Now consider that $B^{\mu}$ is real. The analysis is similar to that
above Eq.~(\ref{eq:m-40-2}), which means we need to consider both
$t$- and $u$-channel diagrams and add a factor of $\frac{1}{2!}$
due to the symmetry of identical particles. Hence the squared amplitude
including the symmetry factor reads:
\begin{eqnarray}
S|{\cal M}|^{2} & = & \frac{|g_{\nu}|^{4}}{2!}\sum_{\epsilon_{1},\thinspace\epsilon_{2}}\sum_{s_{4},\thinspace s_{3}}\left|\overline{u_{4}}P_{L}\left[\slashed{\epsilon}_{2}\frac{i}{\slashed{p}_{F}-m_{F}}\slashed{\epsilon}_{1}+\slashed{\epsilon}_{1}\frac{i}{\slashed{p}'_{F}-m_{F}}\slashed{\epsilon}_{2}\right]P_{R}v_{3}\right|^{2},\label{eq:m-36-1-1}
\end{eqnarray}
where $p_{F}'=p_{2}-p_{3}$ is the momentum of $F$ in the $u$-channel
diagram. 
The remaining calculation is straightforward, though more complicated. 
A convenient approach is to separate the 
summation of vector polarization and the trace of Dirac matrices 
in the way similar to the first step in Eq.~(\ref{eq:m-41}), then 
compute the trace using {\tt Package-X} \cite{Patel:2015tea} before the Lorentz indices are contracted.
The result reads: 
\begin{equation}
S|{\cal M}|^{2}=\frac{|g_{\nu}|^{4}K}{2m_{B}^{4}\left(t-m_{F}^{2}\right){}^{2}\left(u-m_{F}^{2}\right){}^{2}},\label{eq:m-84}
\end{equation}
where
\begin{eqnarray}
K & \equiv & -4m_{B}^{8}\left[6m_{F}^{2}(t+u)-6m_{F}^{4}+t^{2}-8tu+u^{2}\right]\nonumber \\
 &  & -16m_{B}^{6}(t+u)\left(t-m_{F}^{2}\right)\left(u-m_{F}^{2}\right)\nonumber \\
 &  & +m_{B}^{4}\left[m_{F}^{4}\left(7t^{2}-6tu+7u^{2}\right)-8m_{F}^{2}tu(t+u)+4tu\left(t^{2}+u^{2}\right)\right]\nonumber \\
 &  & -4m_{B}^{2}m_{F}^{4}(t-u)^{2}(t+u)+m_{F}^{4}tu(t-u)^{2}.\label{eq:m-85}
\end{eqnarray}
Note that the result is, as it should, symmetric under $t\leftrightarrow u$.

\subsection{$F$ annihilation (scalar case)}

\noindent 
This is the dominant process of $\nu_R$-genesis for subcase (III-2), assuming $B$ is a scalar boson.
In the diagram for subcase (III-2)
in Tab.~\ref{tab:Dominant-processes}, the upper and lower
vertices correspond to $g_{\nu}B\overline{F}P_{R}\nu_{R}$ and $g_{\nu}^{*}B^{\dagger}\overline{\nu_{R}}P_{L}F$. 

As previously discussed {[}see text below Eq.~(\ref{eq:m-34-1}){]},
when $F$ is in the initial state, we treat it as polarized scattering
which implies that we should sum over the initial spins, rather than
taking the average. Thus, the squared amplitude reads:
\begin{equation}
|{\cal M}|^{2}=\sum_{s_{1},\thinspace s_{2}}\sum_{s_{4},\thinspace s_{3}}|g_{\nu}^{*}\overline{u_{4}}P_{L}u_{2}\frac{i}{p_{B}^{2}-m_{B}^{2}}g_{\nu}\overline{v_{1}}P_{R}v_{3}|^{2}.\label{eq:m-40-1}
\end{equation}
The calculation is straightforward and leads to:
\begin{equation}
|{\cal M}|^{2}=\frac{4|g_{\nu}|^{4}}{|t-m_{B}^{2}|^{2}}(p_{1}\cdot p_{3})(p_{2}\cdot p_{4})=|g_{\nu}|^{4}\left(\frac{t-m_{F}^{2}}{t-m_{B}^{2}}\right)^{2}.\label{eq:m-46}
\end{equation}

\subsection{$F$ annihilation (vector case)}

\noindent 
This is the dominant process of $\nu_R$-genesis for subcase (III-2), assuming $B$ is a vector boson.
For a vector mediator, we modify Eq.~(\ref{eq:m-40-1})  as follows:
\begin{eqnarray}
|{\cal M}|^{2} & = & \sum_{s_{1},\thinspace s_{2}}\sum_{s_{4},\thinspace s_{3}}|g_{\nu}^{*}\overline{u_{4}}P_{L}\gamma_{\mu}u_{2}\frac{i}{p_{B}^{2}-m_{B}^{2}}g_{\nu}\overline{v_{1}}\gamma^{\mu}P_{R}v_{3}|^{2}\label{eq:m-40-1-1}\\
 & = & \frac{|g_{\nu}|^{4}}{|p_{B}^{2}-m_{B}^{2}|^{2}}\, {\rm tr \! }\left[\slashed{p}_{4}P_{L}\gamma_{\mu}(\slashed{p}_{2}+m_{2})\gamma_{\nu}P_{R}\right]{\rm tr}\left[(\slashed{p}_{1}-m_{1})\gamma^{\mu}P_{R}\slashed{p}_{3}P_{L}\gamma^{\nu}\right].
\end{eqnarray}
The result is
\begin{equation}
|{\cal M}|^{2}=\frac{16|g_{\nu}|^{4}}{|t-m_{B}^{2}|^{2}}(p_{1}\cdot p_{4})(p_{2}\cdot p_{3})=4|g_{\nu}|^{4}\left(\frac{u-m_{F}^{2}}{t-m_{B}^{2}}\right)^{2}.\label{eq:m-62}
\end{equation}

\section{Approximate estimation\label{sec:Analytic}}

\noindent In this section, we analytically solve Eqs.~(\ref{eq:m-22})
and (\ref{eq:m-23}) with a few crude approximations made on the
collision terms and  the temperature dependence of $g_{\star}^{(\rho)}$
and $g_{\star}^{(P)}$. 

Since $\rho_{{\rm SM}}$ is much larger than
$\rho_{\nu_{R}}$, the energy transfer from SM particles to $\nu_{R}$
has negligible effect on the SM sector. Therefore, the right-hand
side of Eq.~(\ref{eq:m-23}) can be neglected and the co-moving entropy
of the SM sector is conserved, which implies
\begin{equation}
\frac{ds_{{\rm SM}}}{dt}=-3Hs_{{\rm SM}},\label{eq:m-48}
\end{equation}
where $s_{{\rm SM}}$ is the entropy density of the SM. Using Eq.~(\ref{eq:m-48}),
we substitute $dt\rightarrow ds_{{\rm SM}}$ in Eq.~(\ref{eq:m-22})
and obtain 
\begin{equation}
\frac{d\rho_{\nu_{R}}}{ds_{{\rm SM}}}-\frac{4}{3}\frac{\rho_{\nu_{R}}}{s_{{\rm SM}}}\approx-\frac{C_{\nu_{R}}}{3Hs_{{\rm SM}}}.\label{eq:m-29}
\end{equation}
The left-hand side of Eq.~(\ref{eq:m-29}) can be written as a total
derivative according to $d(\rho_{\nu_{R}}s_{{\rm SM}}^{-4/3})=s_{{\rm SM}}^{-4/3}(d\rho_{\nu_{R}}-\frac{4}{3}\rho_{\nu_{R}}s_{{\rm SM}}^{-1}ds_{{\rm SM}})$:
\begin{equation}
\frac{dY}{ds_{{\rm SM}}}\approx-\frac{C_{\nu_{R}}}{3Hs_{{\rm SM}}^{7/3}},\label{eq:m-30}
\end{equation}
where we introduced the yield 
\begin{equation}
Y\equiv\frac{\rho_{\nu_{R}}}{s_{{\rm SM}}^{4/3}}.\label{eq:m-31}
\end{equation}
Therefore, by integrating Eq.~(\ref{eq:m-30}) with respect to $s_{{\rm SM}}$,
we obtain the solution for $Y$: 
\begin{equation}
Y\approx\int_{s_{{\rm SM}}}^{\infty}\frac{C_{\nu_{R}}}{3H\tilde{s}_{{\rm SM}}^{7/3}}d\tilde{s}_{{\rm SM}}. \label{eq:m-32}
\end{equation}
In the freeze-in regime, the contribution of the back-reaction, that is,  
the second part in the squared bracket in Eq.~(\ref{eq:m-21}), is
typically negligible and $C_{\nu_{R}}$ can be approximately treated
as a function of the SM temperature $T$. Since $s_{{\rm SM}}$ is essentially
a function of $T$, for practical use, we write Eq.~(\ref{eq:m-32})
as an integral of $T$: 
\begin{equation}
\rho_{\nu_{R}}(T)\approx s_{{\rm SM}}^{4/3}(T)\int_{T}^{\infty}\frac{C_{\nu_{R}}(\tilde{T})}{3H(\tilde{T})s_{{\rm SM}}^{7/3}(\tilde{T})}s_{{\rm SM}}'(\tilde{T})d\tilde{T}.\label{eq:m-49}
\end{equation}
Eq.~(\ref{eq:m-49}) is the formula we will use to approximately
estimate the abundance of $\nu_{R}$. To proceed with the integration
in Eq.~(\ref{eq:m-49}), we need to take some power-law approximations.

\subsection{Power-law approximation of  collision terms}

\subsubsection*{Decay processes}

\noindent For decay processes, when the contribution of back-reaction
can be neglected, we estimate the collision term as follows 
\begin{equation}
C_{\nu_{R}}\sim N_{\nu_{R}}S|{\cal M}|^{2}\int E_{3}d\Pi_{1}d\Pi_{3}d\Pi_{4}(2\pi)^{4}\delta^{4}(p_{1}-p_{3}-p_{4})f_{1},\label{eq:m-50}
\end{equation}
where $S|{\cal M}|^{2}$  for decay processes is actually a constant
that can be fully determined by  $m_{B}$, $m_{F}$ and $g_{\nu}$---see
Tab.~\ref{tab:Dominant-processes}. Therefore, we can extract it
out of the integral. The $\delta$ function can be removed using the
procedure introduced in Appendix.~\ref{sec:Numerical-collision}.
According to Eq.~(\ref{eq:m-2-1}), we get
\begin{equation}
C_{\nu_{R}}\sim N_{\nu_{R}}S|{\cal M}|^{2}\int E_{3}\frac{|\boldsymbol{p}_{1}|^{2}d|\boldsymbol{p}_{1}|dc_{1}d\phi_{1}}{(2\pi)^{3}2E_{1}}\frac{|\boldsymbol{p}_{3}|^{2}dc_{3}d\phi_{3}}{(2\pi)^{3}2E_{3}}\frac{2\pi}{2E_{4}}J^{-1}f_{1},\label{eq:m-51}
\end{equation}
where $dc_i = d \cos \theta_i$ and $J$ is an ${\cal O}(1)$ quantity with its explicit form given
in Eq.~(\ref{eq:m-12}). We further make the approximation that
$f_{1}$ is either ${\cal O}(1)$ or exponentially suppressed, for
$T>E_{1}/3$ or $T<E_{1}/3$, respectively. Therefore, we can remove
$J^{-1}f_{1}$ in Eq.~(\ref{eq:m-51}) and replace $\int d|\boldsymbol{p}_{i}|\rightarrow T$:
\begin{equation}
C_{\nu_{R}}\sim N_{\nu_{R}}S|{\cal M}|^{2}\frac{4\pi\langle|\boldsymbol{p}_{1}|^{2}\rangle T}{(2\pi)^{3}2\langle E_{1}\rangle}\frac{4\pi\langle|\boldsymbol{p}_{3}|^{2}\rangle}{2(2\pi)^{3}}\frac{2\pi}{2\langle E_{4}\rangle}.\label{eq:m-52}
\end{equation}
Here $4\pi$ comes from $\int dc_{i}d\phi_{i}$ and ``$\langle\ \rangle$'' stands for mean values in the integral. Note that when $T\ll m_{1}$,
$f_{1}$ would exponentially suppress the result. So we only consider 
the regime in which the temperature is larger or comparable to $m_{1}$,
which implies that  $\langle E_{i}\rangle$ and $\langle|\boldsymbol{p}_{i}|^{2}\rangle$
are roughly of the order of $T$ and $T^{2}$. Hence we replace $\langle E_{i}\rangle\rightarrow T$,
$\langle|\boldsymbol{p}_{i}|^{2}\rangle\rightarrow T^{2}$ and get
\begin{equation}
C_{\nu_{R}}\sim\begin{cases}
\frac{1}{16\pi^{3}}N_{\nu_{R}}S|{\cal M}|^{2}T^{3} & \ \ (T\gtrsim m_{1}/3)\\
0 & \ \ (T\lesssim m_{1}/3)
\end{cases},\ ({\rm for\ }B/F\ {\rm decay})\thinspace,\label{eq:m-53}
\end{equation}
where $m_{1}$ is $m_{B}$ or $m_{F}$ if the initial particle
is $B$ or $F$, respectively.

\subsubsection*{Annihilation processes}

\noindent For annihilation processes, the derivation is similar though  there are two noteworthy differences. First, there is an additional
$\langle d\Pi_{2}\rangle\sim\frac{4\pi\langle|\boldsymbol{p}_{2}|^{2}\rangle T}{(2\pi)^{3}2\langle E_{2}\rangle}$,
which contributes to $C_{\nu_{R}}$ by a factor of $\frac{T^{2}}{(2\pi)^{2}}$.
Besides, since $S|{\cal M}|^{2}$ depends on the momenta in the integral,
to extract it out of the integral we replace it with its mean value
and obtain
\begin{equation}
C_{\nu_{R}}\sim\begin{cases}
\frac{1}{64\pi^{5}}N_{\nu_{R}}\langle S|{\cal M}|^{2}\rangle T^{5} & \ \ (T\gtrsim m_{1}/3)\\
0 & \ \ (T\lesssim m_{1}/3)
\end{cases},\ ({\rm for\ }B/F\ {\rm annihilation}),\label{eq:m-54}
\end{equation}
where $m_1$ is $m_{B}$ or $m_{F}$, depending on which particles annihilates.
To estimate $\langle S|{\cal M}|^{2}\rangle$, we neglect some ${\cal O}(1)$
quantities in the expressions in Tab.~\ref{tab:Dominant-processes}
and take  $t\rightarrow-2\langle p_{1}\cdot p_{3}\rangle\sim-2T^{2}$,
$u\rightarrow-2\langle p_{1}\cdot p_{4}\rangle\sim-2T^{2}$. The result
reads
\begin{equation}
\langle S|{\cal M}|^{2}\rangle\sim|g_{\nu}|^{4}\frac{T^{4}}{(T^{2}+m_{X}^{2}/2)^{2}},\label{eq:m-64}
\end{equation}
where $m_{X}$ denotes the mediator mass:
\begin{equation}
m_{X}\equiv\begin{cases}
m_{F} & \ \textrm{for\ case\ (II-2)}\\
m_{B} & \ \textrm{for\ case\ (III-2)}
\end{cases}.\label{eq:m-70}
\end{equation}
Substituting Eq.~(\ref{eq:m-64}) in Eq.~(\ref{eq:m-54}), we obtain
\begin{equation}
C_{\nu_{R}}\sim\begin{cases}
\frac{1}{64\pi^{5}}N_{\nu_{R}}|g_{\nu}|^{4}T^{5} & \frac{m_{X}}{\sqrt{2}}\lesssim T\\[2mm]
\frac{1}{16\pi^{5}}N_{\nu_{R}}|g_{\nu}|^{4}m_{X}^{-4}T^{9} & \frac{1}{3}m_{1}\lesssim T\lesssim\frac{m_{X}}{\sqrt{2}}\\[2mm]
0 & T\lesssim\frac{1}{3}m_{1}
\end{cases},\ ({\rm for\ }B/F\ {\rm annihilation}),\label{eq:m-65}
\end{equation}
where $m_{X}$ is defined in Eq.~(\ref{eq:m-70}), $m_{1}$ takes
$m_{B}$ for subcase (II-2) or $m_{F}$ for subcase (III-2), respectively. 

\begin{figure}
\centering

\includegraphics[width=0.7\textwidth]{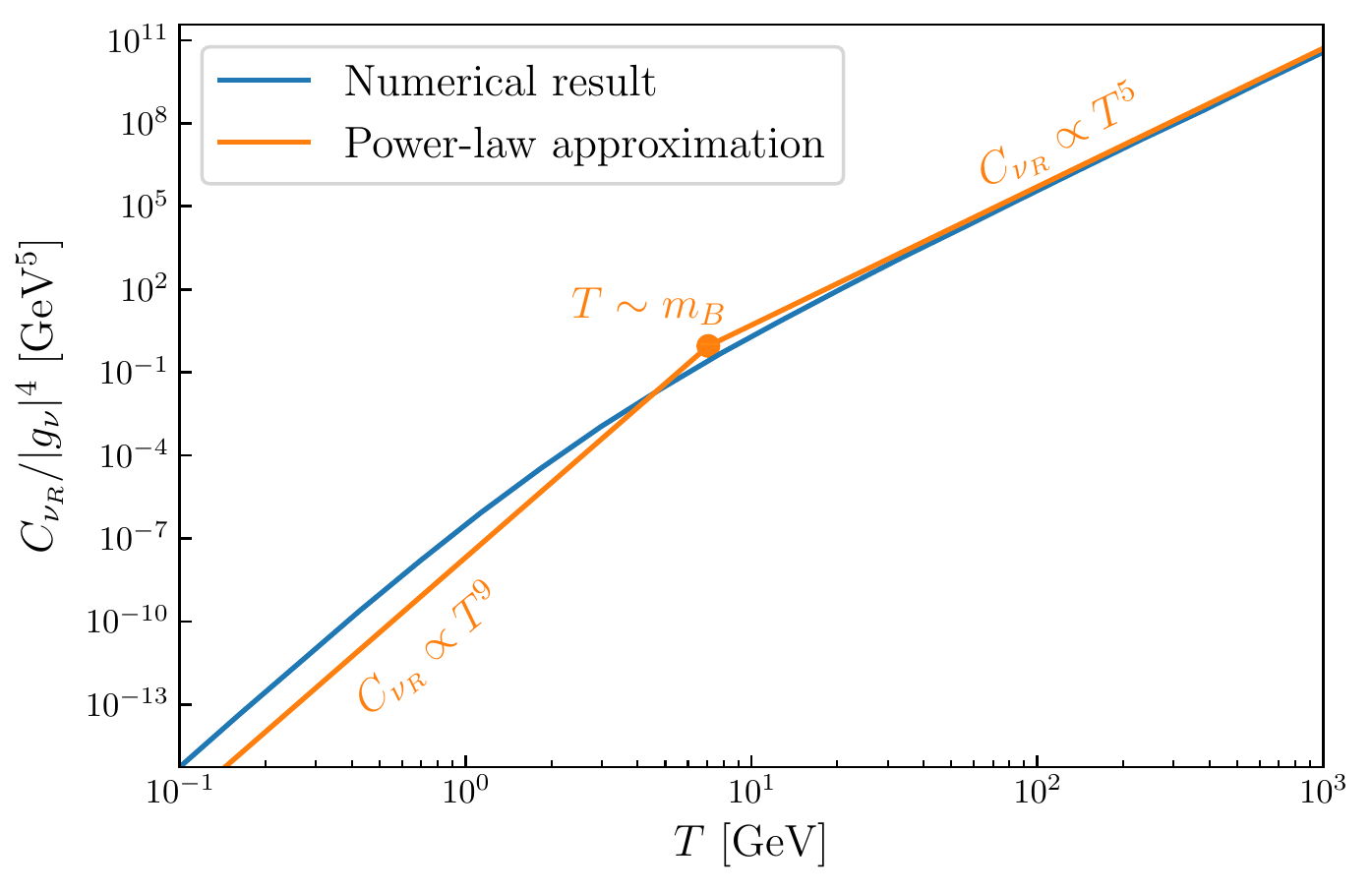}

\caption{Power-law approximation of the collision term of $F+\overline{F}\rightarrow\nu_{R}+\overline{\nu_{R}}$
compared with the numerical (exact) result. In this illustration,
the approximate curve is produced according to Eq.~(\ref{eq:m-65})
with $N_{\nu_{R}}=1$, $m_{B}=10$ GeV and $m_{F}=0.1$ GeV. The numerical
result is obtained using the method in Appendix~\ref{sec:Numerical-collision}
with the same values, assuming $B$ is a scalar and initial/final
particles obey  Fermi-Dirac statistics. \label{fig:Power-law-approximation} 
}
\end{figure}

Eqs.~(\ref{eq:m-53}) and (\ref{eq:m-65}) are our power-law 
approximations of  collision terms for decay and annihilation processes,  respectively.
Since we have used several  approximations in the derivation, it
should only be an estimation of the order of magnitude. In Fig.~\ref{fig:Power-law-approximation},
we compare our power-law approximation of the collision term for subcase
(III-2) with the exact result which is obtained using the method introduced
in Appendix~\ref{sec:Numerical-collision}.

\subsection{Approximate result}

\noindent With the power-law approximations of collision terms in
Eqs.~(\ref{eq:m-53}) and (\ref{eq:m-65}), we are ready to approximately
estimate the abundance of $\nu_{R}$ using the integral in Eq.~(\ref{eq:m-49}).
The Hubble parameter is determined by $H^{2}=8\pi\rho_{{\rm tot}}/(3m_{{\rm pl}}^{2})$, 
where $\rho_{{\rm tot}}$ is the total energy density and $m_{{\rm pl}}=1.22\times10^{19}$
GeV is the Planck mass. We take $\rho_{{\rm tot}}\approx\rho_{{\rm SM}}$
in the Hubble parameter so that
\begin{equation}
H\approx\sqrt{\frac{8\pi^{3}g_{\star}^{(\rho)}}{90}}\frac{T^{2}}{m_{{\rm pl}}}.\label{eq:m-28}
\end{equation}
In the SM entropy density,
\begin{equation}
s_{{\rm SM}}(T)=\frac{2\pi^{2}}{45}g_{\star}^{(s)}T^{3},\label{eq:m-66}
\end{equation}
we neglect the small difference between $g_{\star}^{(s)}$ and $g_{\star}^{(\rho)}$,
and use $g_{\star}\approx g_{\star}^{(s)}\approx g_{\star}^{(\rho)}$.
In addition, we treat $g_{\star}$ as a constant inside the integral.
When we compute the derivative $s_{{\rm SM}}'(T)$ and the integral,
the mean value $\langle g_{\star}\rangle$ is used instead of $g_{\star}$.

For the following power-law form of $C_{\nu_{R}}$, 
\begin{equation}
C_{\nu_{R}}(T)\approx\Lambda^{n-1}T^{6-n},\ (n>0),\label{eq:m-67}
\end{equation}
the integral in Eq.~(\ref{eq:m-49}) converges for $T\rightarrow\infty$.
This can be seen from power counting: $s_{{\rm SM}}\sim T^{3}$, $s_{{\rm SM}}'\sim T^{2}$,
$H\sim T^{2}$, $C_{\nu_{R}}s_{{\rm SM}}'/(Hs_{{\rm SM}}^{7/3})\sim1/T^{n+1}$.
To make the integral $\int^{\infty}\frac{1}{T^{n+1}}dT$ converge 
for  $T\rightarrow\infty$, we need $n>0$. Therefore, in the freeze-in
mechanism when $T$ increases to sufficiently large vales, $C_{\nu_{R}}$
should increase slower than $T^{6}$. Indeed,  one can see that both
Eqs.~(\ref{eq:m-53}) and (\ref{eq:m-65}) satisfy this requirement. 

Substituting Eqs.~(\ref{eq:m-53}) and (\ref{eq:m-65}-\ref{eq:m-66}) in Eq.~(\ref{eq:m-49}), we obtain 
\begin{equation}
\frac{\rho_{\nu_{R}}}{\rho_{{\rm SM}}}\sim N_{\nu_{R}}S|{\cal M}|^{2}\frac{15\sqrt{5}g_{\star}^{1/3}m_{\text{pl}}}{16\pi^{13/2}\langle g_{\star}\rangle^{11/6}}\times\begin{cases}
T^{-3} & \ \ (T\gtrsim m_{1}/3)\\
(m_{1}/3)^{-3} & \ \ (T\lesssim m_{1}/3)
\end{cases},\ {\rm for\ }B/F\ {\rm decay},\label{eq:m-68}
\end{equation}
and 
\begin{equation}
\frac{\rho_{\nu_{R}}}{\rho_{{\rm SM}}}\sim N_{\nu_{R}}|g_{\nu}|^{4}\frac{45\sqrt{5}g_{\star}^{1/3}m_{\text{pl}}}{64\pi^{17/2}\langle g_{\star}\rangle^{11/6}}\begin{cases}
\frac{1}{T} & \ \frac{m_{X}}{\sqrt{2}}\lesssim T\\[2mm]
\frac{4\sqrt{2}}{3m_{X}}-\frac{4T^{3}}{3m_{X}^{4}} & \ \frac{m_{1}}{3}\lesssim T\lesssim\frac{m_{X}}{\sqrt{2}}\\[2mm]
\frac{4\sqrt{2}}{3m_{X}}-\frac{4m_{1}^{3}}{81m_{X}^{4}} & \ T\lesssim\frac{m_{1}}{3}
\end{cases},\ {\rm for\ }B/F\ {\rm annihilation}.\label{eq:m-69}
\end{equation}
Note that $g_{\star}=g_{\star}(T)$ is a $T$-dependent quantity
and $\langle g_{\star}\rangle$ is the effective mean value used in
the integral. As an approximation, one can take $\langle g_{\star}\rangle\sim g_{\star}(T=m_{X})$
in Eq.~(\ref{eq:m-69}) or $\langle g_{\star}\rangle\sim g_{\star}(T=m_{1})$
in Eq.~(\ref{eq:m-68}),  because $\nu_{R}$ is the most efficiently
produced at this temperature. 

We further translate the results of $\rho_{\nu_{R}}/\rho_{{\rm SM}}$
into $\Delta N_{{\rm eff}}$ according to Eq.~(\ref{eq:m-87}), which results in  
\begin{equation}
\Delta N_{{\rm eff}}\sim2.7\frac{m_{\text{pl}}S|{\cal M}|^{2}}{\langle g_{\star}\rangle^{11/6}m_{1}^{3}}\sim0.1\times\left(\frac{100}{\langle g_{\star}\rangle}\right)^{11/6}\left(\frac{700\ {\rm GeV}}{m_{1}}\right)\left|\frac{g_{\nu}}{10^{-7}}\right|^{2},\label{eq:m-89}
\end{equation}
for $B$ or $F$ decay, and
\begin{equation}
\Delta N_{{\rm eff}}\sim1.4\times10^{-2}\frac{m_{\text{pl}}|g_{\nu}|^{4}}{\langle g_{\star}\rangle^{11/6}m_{X}}\sim0.1\times\left(\frac{100}{\langle g_{\star}\rangle}\right)^{11/6}\left(\frac{400\ {\rm GeV}}{m_{X}}\right)\left|\frac{g_{\nu}}{10^{-3}}\right|^{4},\label{eq:m-90}
\end{equation}
for $B$ or $F$ annihilation.

Eqs.~(\ref{eq:m-89}) and (\ref{eq:m-90}) are our final results for the 
approximate estimation. Here $m_{1}$ is the initial particle mass
and $m_{X}$ is $m_F$ for case (II-2) and $m_B$ for case (III-2). We stress  that
the results presented here are based on several approximations 
which might deviate from the exact result by one or even two orders
of magnitude---see Fig.~\ref{fig:Power-law-approximation} for example.
The results should only be used to qualitatively estimate the order
of magnitude. In particular, since we ignored the back-reaction, it
would be incorrect to apply Eqs.~(\ref{eq:m-89}) and (\ref{eq:m-90})
to large $\Delta N_{{\rm eff}}$ due to saturated production rates.
If the freeze-in process happens at temperatures well above the electroweak
scale and $\nu_{R}$ has been decoupled since then, we know that $\Delta N_{{\rm eff}}$
should be smaller than 0.14~\cite{Abazajian:2019oqj,Luo:2020sho}. This  provides a useful
criterion to check whether the back-reaction can be neglected or not.

In the next section, we will discuss an example in which our approximate
result is compared with the exact one, namely when Dirac neutrino masses are generated by the SM Higgs mechanism.

\section{The SM Higgs as an example\label{sec:Higgs}}

\noindent Let us assume neutrinos are Dirac particles and their masses originate
from tiny Yukawa couplings with the SM Higgs (flavor indices are ignored here), 
\begin{equation}
{\cal L}\supset Y_{\nu}\overline{L}\tilde{H}\nu_{R},\label{eq:m-76}
\end{equation}
where $L=(\nu_{L},\ e_{L})^{T}$, $\tilde{H}=i\sigma_{2}H^{*}$ and
$H=\frac{1}{\sqrt{2}}(0,v+h)^{T}$ in the unitary gauge. 
Here $h$ is the Higgs boson and $v\approx246$ GeV. Eq.~(\ref{eq:m-76}) gives
rise to neutrino masses $m_{\nu}=\frac{v}{\sqrt{2}}Y_{\nu}$, which
implies that the Yukawa couplings should be 
\begin{equation}
Y_{\nu}=\sqrt{2}\frac{m_{\nu}}{v}=5.7\times10^{-13}\left(\frac{m_{\nu}}{0.1\ {\rm eV}}\right).\label{eq:m-47}
\end{equation}
In the unitary gauge $\nu_{R}$ couples to the SM only via ${\cal L}\supset\frac{Y_{\nu}}{\sqrt{2}}h\overline{\nu_{L}}\nu_{R}$.
 According to our discussion in Sec.~\ref{sec:basic}, the dominant
process\footnote{At low temperatures ($T\ll m_{h}$), other processes such as $\nu_{L}+\overline{\nu}_{L}\rightarrow\nu_{R}+\overline{\nu}_{R}$
have higher production rates than $h\rightarrow\nu_{R}+\overline{\nu}_{L}$
because the latter is exponentially suppressed. However, the overall
contribution of the former to the accumulated $\rho_{\nu_{R}}$ is
still negligible, which can be estimated using the power-law approximation
in Sec.~\ref{sec:Analytic}. } for $\nu_{R}$ production is Higgs decay: $h\rightarrow\nu_{R}+\overline{\nu}_{L}$.
According to Tab.~\ref{tab:Dominant-processes}, the squared amplitude
is
\begin{equation}
|{\cal M}|^{2}=\frac{1}{2}Y_{\nu}^{2}m_{h}^{2},\label{eq:m-77}
\end{equation}
where $m_{h}\approx125$ GeV is the Higgs mass. In the Maxwell-Boltzmann
(MB) approximation, the collision term of $h\rightarrow\nu_{R}+\overline{\nu}_{L}$
can be computed analytically according to Appendix~\ref{sec:3particle}---see
also Ref.~\cite{Escudero:2020dfa}. The result reads:
\begin{equation}
C_{\nu_{R}}\approx N_{\nu_{R}}|{\cal M}|^{2}\frac{m_{h}^{2}}{64\pi^{3}}TK_{2}\left(\frac{m_{h}}{T}\right),\ \ {\rm (MB\ approximation}),\label{eq:m-71}
\end{equation}
where $K_{2}$ is a $K$-type Bessel function  of order $2$. Since
$K_{2}(x)\approx2x^{-2}$ for $x\ll1$ and $K_{2}(x)\sim e^{-x}$
for $x\gg1$, Eq.~(\ref{eq:m-71}) is approximately consistent with
the power-law approximation in Eq.~(\ref{eq:m-53}).

To obtain the exact result using Bose-Einstein and Fermi-Dirac distributions,
one has to invoke Monte-Carlo integration, which is detailed in Appendix~\ref{sec:Numerical-collision}.
In Fig.~\ref{fig:SM_Higgs}, we present the results obtained from
exact numerical calculations and the aforementioned approximations
(MB and power-law).

\begin{figure}
\centering

\includegraphics[width=0.7\textwidth]{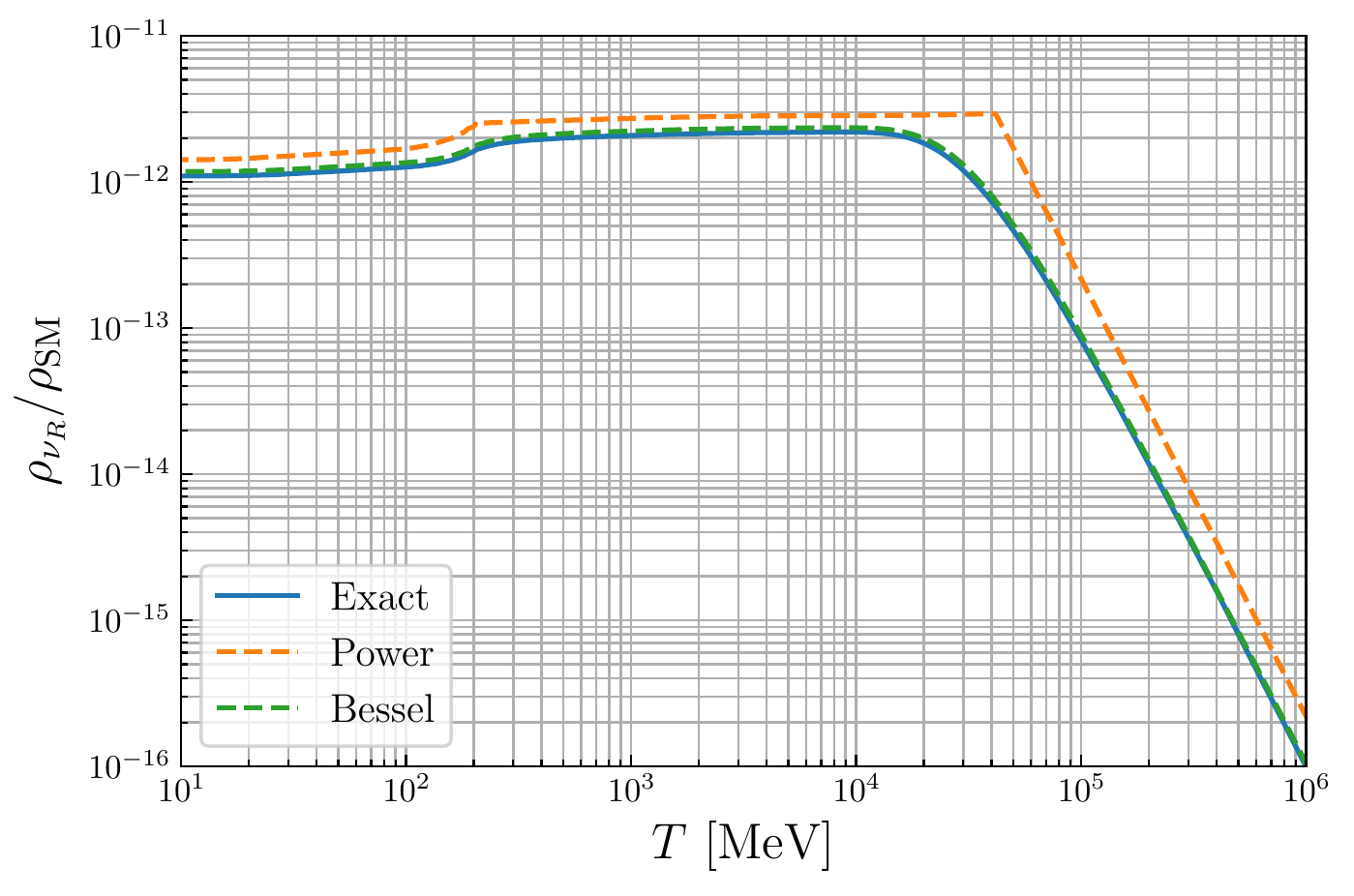}

\caption{The SM Higgs as an example. Taking the Yukawa coupling in Eq.~(\ref{eq:m-47}),
we compute the effect of the Higgs-$\nu_{R}$-$\nu_{L}$ coupling on the $\nu_{R}$
abundance in the early Universe and obtain $\Delta N_{{\rm eff}}\approx7.5\times10^{-12}$.
The blue curve is obtained by numerically solving the Boltzmann equation
and invoking Monte-Carlo integration of the phase space. The orange
curve is obtained using the power-law approximation---see Eq.~(\ref{eq:m-68}).
The green curve assumes Maxwell-Boltzmann statistics, so that the
collision term can be analytically formulated as a Bessel function
in Eq.~(\ref{eq:m-71}). \label{fig:SM_Higgs} }
\end{figure}

Taking the low-temperature value of the blue curve in Fig.~\ref{fig:SM_Higgs}
and using Eq.~(\ref{eq:m-87}), we obtain
\begin{equation}
    \Delta N_{{\rm eff}}\approx7.5\times10^{-12}\left(\frac{m_{\nu}}{0.1\ {\rm eV}}\right)^{2}. 
\end{equation}
This is a precise result on $\Delta N_{{\rm eff}}$ that originates 
from the SM Higgs interaction with Dirac neutrinos. 

\section{Numerical approach\label{sec:Numerical-results}}

\noindent In this section, we numerically solve
\footnote{The code is publicly available at \url{https://github.com/xuhengluo/Thermal_Boltzmann_Solver}.} the Boltzmann equations 
(\ref{eq:m-22}) and (\ref{eq:m-23}) to investigate the evolution
of the $\nu_{R}$ abundance for all cases outlined in Tab.\ \ref{tab:Dominant-processes}. 
Although solving the differential equation
itself is not difficult, computing the collision term $C_{\nu_{R}}$
which is a 9- or 12-dimensional integral is computationally expensive.
In some simple cases, the collision term is analytically calculable
assuming that all thermal species obey the Maxwell-Boltzmann statistics.
Known examples include decay of a massive particle to two massless
particles (used in Section \ref{sec:Higgs}) and $2\rightarrow2$ scattering of four massless particles
with contact interactions. The analytical expressions can be derived
following the calculations in Appendix A of Ref.~\cite{Dolgov:1997mb} and Appendix D of Ref.~\cite{Fradette:2018hhl},
and the results can be found, e.g., in Appendix~\ref{sec:3particle}
of this paper (for $1\rightarrow2$) or Appendix~C in Ref.~\cite{Luo:2020sho}
(for $2\rightarrow2$). More complicated collision terms with Fermi-Dirac/Bose-Einstein statistics and/or with more massive
states and/or with and non-contact interactions, 
can only be evaluated accurately via numerical approaches.

For numerical evaluation of high-dimensional integrals, usually one
has to adopt the Monte-Carlo method. Monte-Carlo integration of multi-particle
phase space is often used in collider phenomenology studies and has
been implemented in a variety of packages including {\tt CalcHEP}~\cite{Belyaev:2012qa}
and similar other  tools. However, since the Monte-Carlo module in
{\tt CalcHEP} is more dedicated to calculations of cross sections,
in order to compute the collision terms more conveniently and efficiently\footnote{In a thermal distribution, the particle energy in principle can be
infinitely large, though this  is exponentially suppressed. To improve
the efficiency of computation, we  include this property of collision
terms directly in the Monte-Carlo module.}, we develop our own Monte-Carlo module
using similar techniques to that in Appendix I of the {\tt CalcHEP}
manual\footnote{See \url{http://theory.npi.msu.su/~pukhov/CALCHEP/calchep_man_3.3.6.pdf}}.
The details are presented in Appendix~\ref{sec:Numerical-collision}.
As aforementioned, for both $1\rightarrow2$ and $2\rightarrow2$
processes, there are special cases with known analytical results.
We have checked that our Monte-Carlo module can accurately reproduce
those results.

It it important to note  that when the $\nu_{R}$ temperature $T_{\nu_{R}}$
is much smaller than the SM temperature $T$, the collision term
$C_{\nu_{R}}(T,\ T_{\nu_{R}})$, as a function of $T$ and $T_{\nu_{R}}$,
is almost exclusively determined by $T$, i.e.,  $C_{\nu_{R}}(T,\ T_{\nu_{R}})\approx C_{\nu_{R}}(T,\ 0)$.
 When $T_{\nu_{R}}$ is approaching $T$, in order to take the back-reaction
into account, we use
\begin{equation}
C_{\nu_{R}}(T,\ T_{\nu_{R}})\approx C_{\nu_{R}}(T,\ 0)-C_{\nu_{R}}(T_{\nu_{R}},\ 0),\label{eq:m-98}
\end{equation}
which, as we have numerically checked, turns out to be a rather accurate
approximation.  Note that $C_{\nu_{R}}(T,\ T_{\nu_{R}})$ constructed
in this way satisfies the condition of thermal equilibrium: $C_{\nu_{R}}(T,\ T_{\nu_{R}})=0$
when $T=T_{\nu_{R}}$. Furthermore, this treatment can be justified
from analytical results as well. Taking subcase (III-2) for example,
we know that when $m_{F}\ll T\ll m_{B}$ there
is an analytical result: $C_{\nu_{R}}\propto T^{9}-T_{\nu_{R}}^{9}$~\cite{Luo:2020sho},
which indeed can be decomposed in the form of Eq.~(\ref{eq:m-98}).

We comment here that when $\nu_{R}$ is not in thermal equilibrium,
the temperature $T_{\nu_{R}}$ is not well defined. Actually particles
produced by freeze-in usually have non-thermal distributions very
different from the Fermi-Dirac one (see e.g.\ \cite{Bae:2017dpt,Ballesteros:2020adh}). Nevertheless, we find that in our case using the Fermi-Dirac
distribution for $\nu_{R}$ causes very little deviation from the
true value because  
the shapes of $f_{3}$ and $f_{4}$ affect the result mainly via the
backreaction term which is negligible when $\rho_{\nu_{R}}$ is
small. When $\rho_{\nu_{R}}$ saturates the upper bound of thermal
equilibrium, it enters the freeze-out regime where the Fermi-Dirac
distribution with a well-defined $T_{\nu_{R}}$ can be used. Only
in a quite narrow window when $\rho_{\nu_{R}}/\rho_{\nu_{L}}$ is
approaching $1$ (i.e.\ in the transition from the freeze-in to freeze-out
regimes), the specific form of backreaction matters. 
We leave possible refinements in this window to future work.

By applying the Monte-Carlo procedure to each process in Tab.~\ref{tab:Dominant-processes}
with the assumption of Eq.~(\ref{eq:m-98}), we obtain the numerical
values of the collision terms which will be passed to the differential
equation solver to solve $\rho_{\nu_{R}}$. Theoretically, the Boltzmann
equations should be solved starting from the initial point at $T=\infty$
with $\rho_{\nu_{R}}=0$. According to our power-law analyses in Sec.~\ref{sec:Analytic},
if we set the initial point at a finite $T$ with $\rho_{\nu_{R}}=0$,
the deviation $\delta\rho_{\nu_{R}}$ from the true value is
\begin{equation}
\delta\rho_{\nu_{R}}/\rho_{\nu_{R}}\sim\begin{cases}
{\cal O}(m_{B,F}^{3}/T^{3}) & {\rm \ for\ decay}\\
{\cal O}(m_{B,F}/T) & {\rm \ for\ annihilation}
\end{cases},\label{eq:m-99}
\end{equation}
where $m_{B,F}=\max(m_B,\ m_F)$.
Therefore to limit the error within, e.g., $1\%$, one only needs
to set $T>{\cal O}(10^{2}\, m_{B,F})$. 

Last, we note that the Boltzmann equations (\ref{eq:m-22}) and (\ref{eq:m-23})
can be combined as
\begin{equation}
\frac{d\rho_{\nu_{R}}}{d\rho_{{\rm SM}}}=\frac{4H\rho_{\nu_{R}}-C_{\nu_{R}}^{(\rho)}}{3H(\rho_{{\rm SM}}+P_{{\rm SM}})+C_{\nu_{R}}^{(\rho)}}.\label{eq:m-91}
\end{equation}
We use Eq.~(\ref{eq:m-91}) to avoid involving the time parameter
$t$ for the sake of  stability of the Boltzmann equation solver. 
Occasionally (when $\nu_R$ is strongly coupled to the SM plasma),
we use ${dT_{\nu_{R}}}/{dT_{{\rm SM}}}$  instead of ${d\rho_{\nu_{R}}}/{d\rho_{{\rm SM}}}$ and impose an upper bound
$T_{\nu_{R}}\leq T_{\rm SM}$ in the Boltzmann equation solver. \\

\begin{figure}
    \centering
    
    \includegraphics[width=0.48\textwidth]{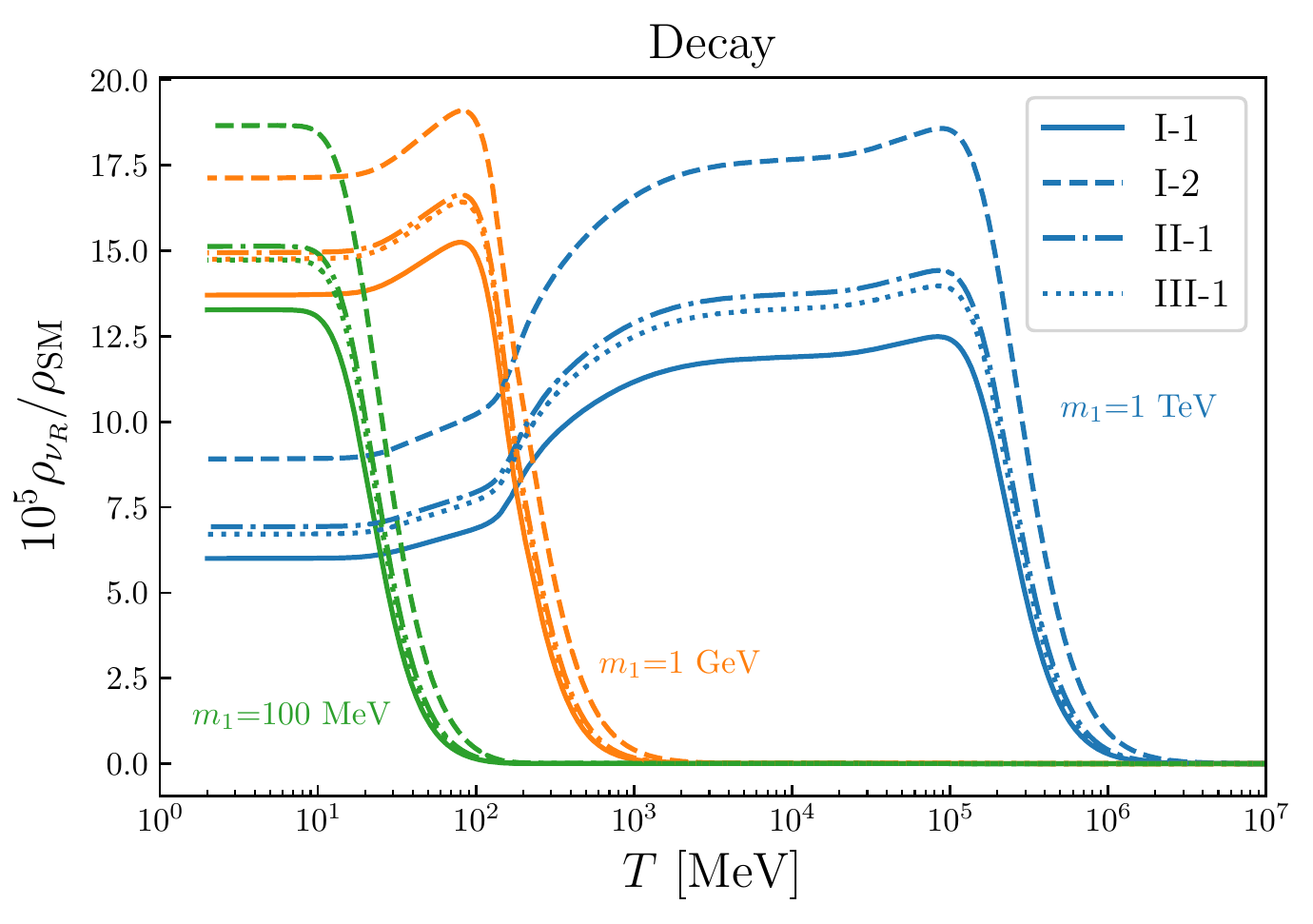}\includegraphics[width=0.48\textwidth]{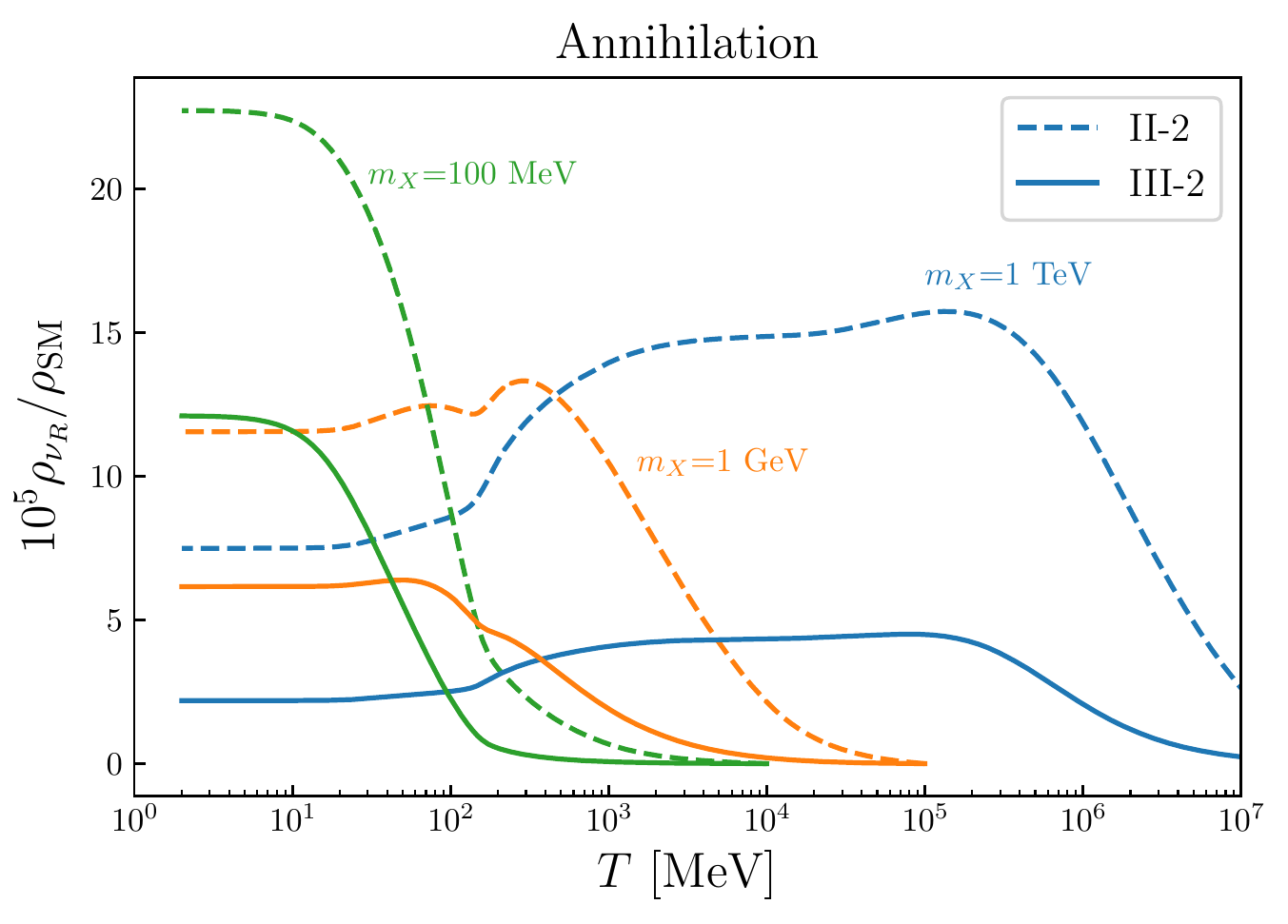}
    
    \includegraphics[width=0.48\textwidth]{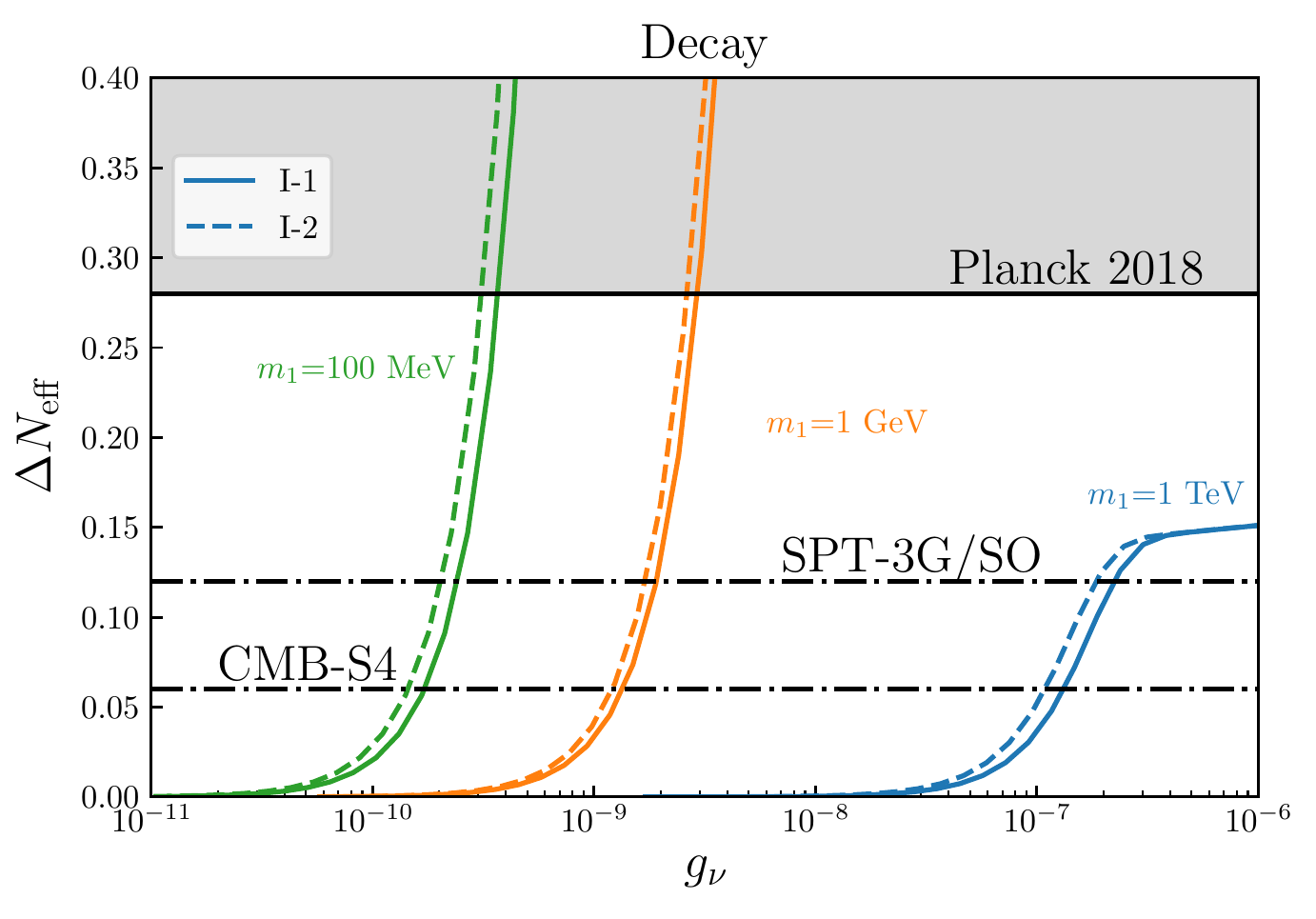}\includegraphics[width=0.48\textwidth]{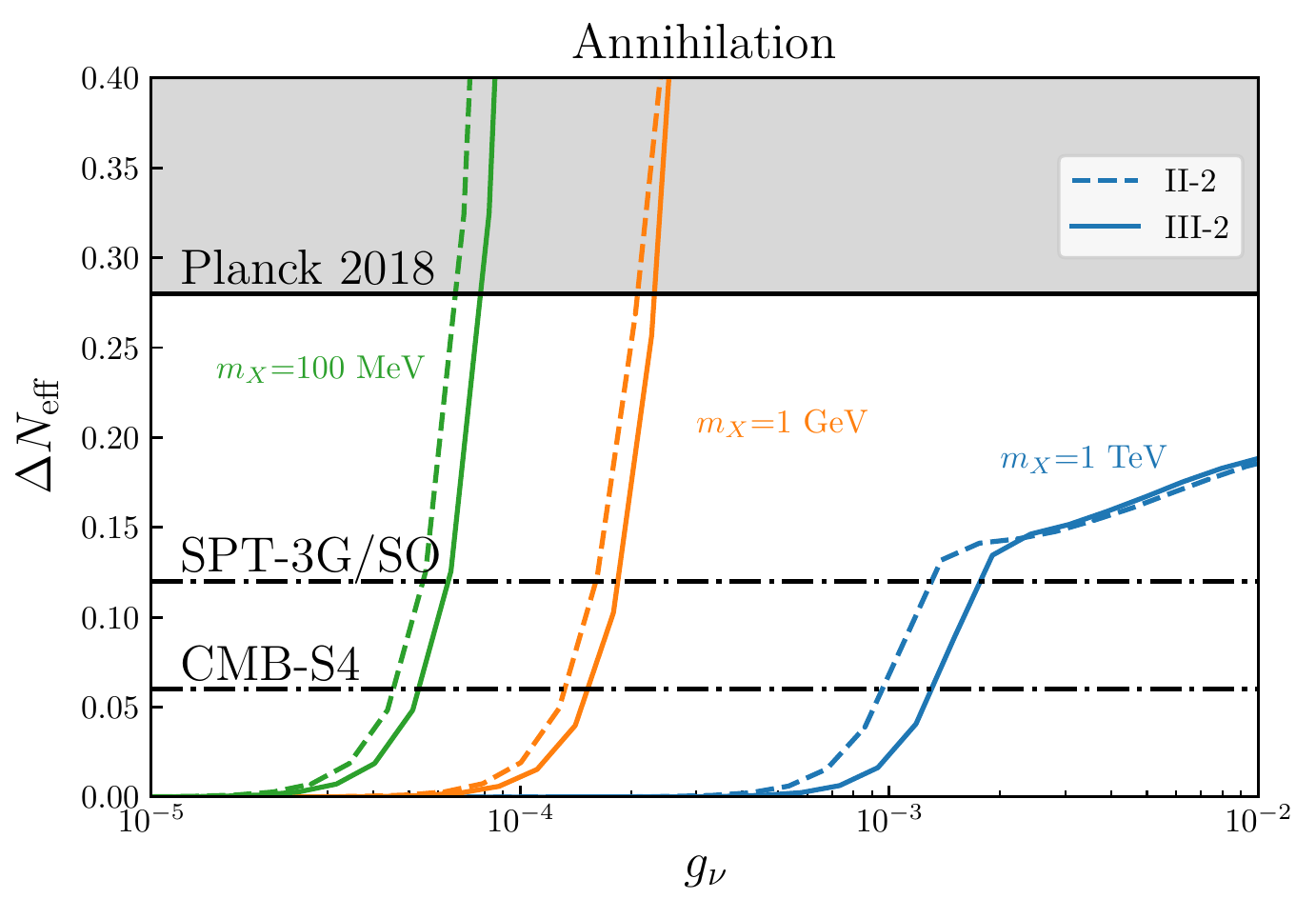}
    
    \caption{Upper panels: the energy density of right-handed neutrinos $\rho_{\nu_{R}}$
    obtained by numerically solving the Boltzmann equations (\ref{eq:m-22})
    and (\ref{eq:m-23}) for all  cases listed in Tab.~\ref{tab:Dominant-processes}.
    Lower panels: contributions of $\nu_{R}$ to $N_{{\rm eff}}$ for
    varying $g_{\nu}$. Here $m_1$ is the initial particle mass of the decay process 
    and $m_X$ is the internal propagator mass of the annihilation process.
    Other relevant parameters are specified in the text. \label{fig:N_eff_g_decay}}
    
\end{figure}

\begin{figure}
    \centering
    
    \includegraphics[width=0.48\textwidth]{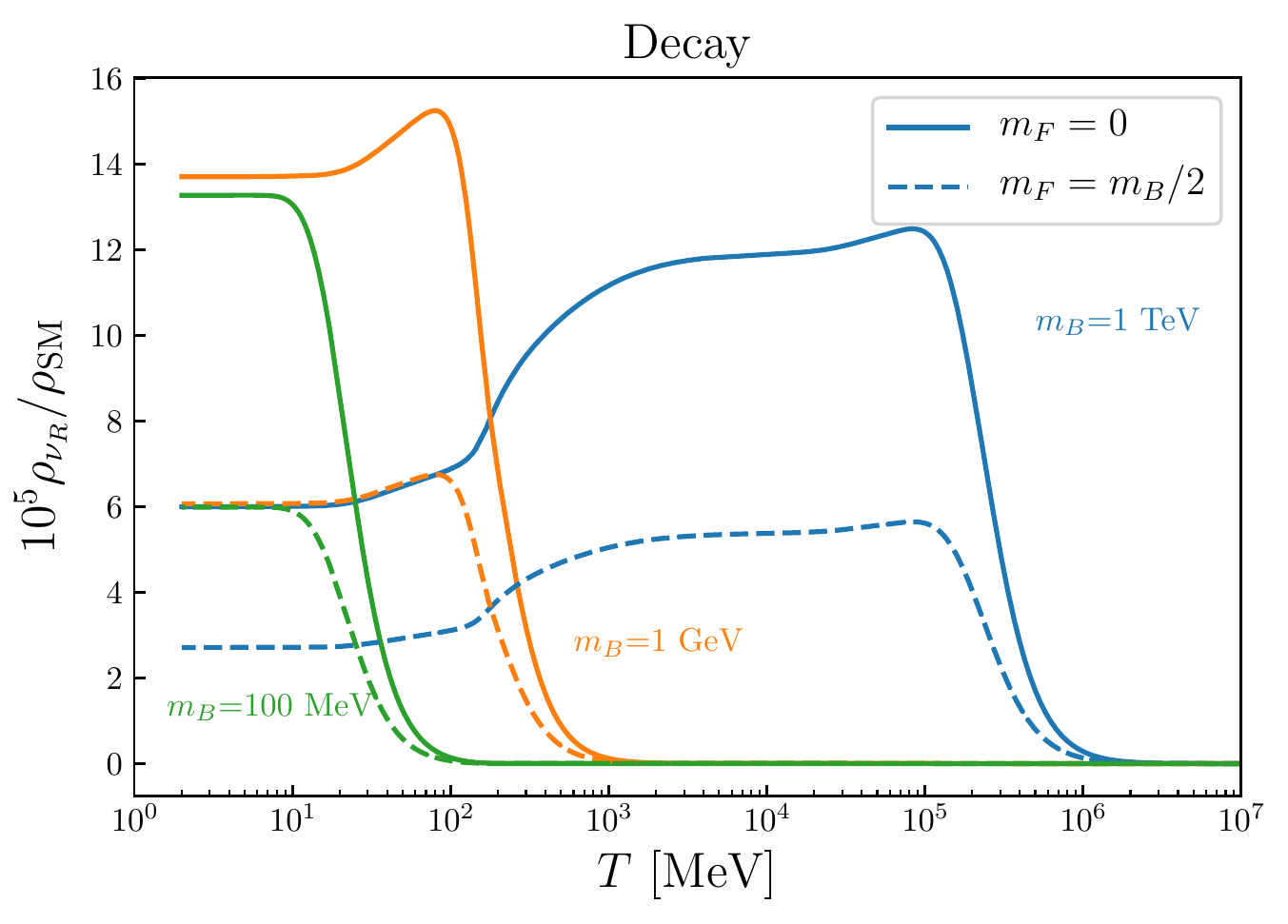}\includegraphics[width=0.48\textwidth]{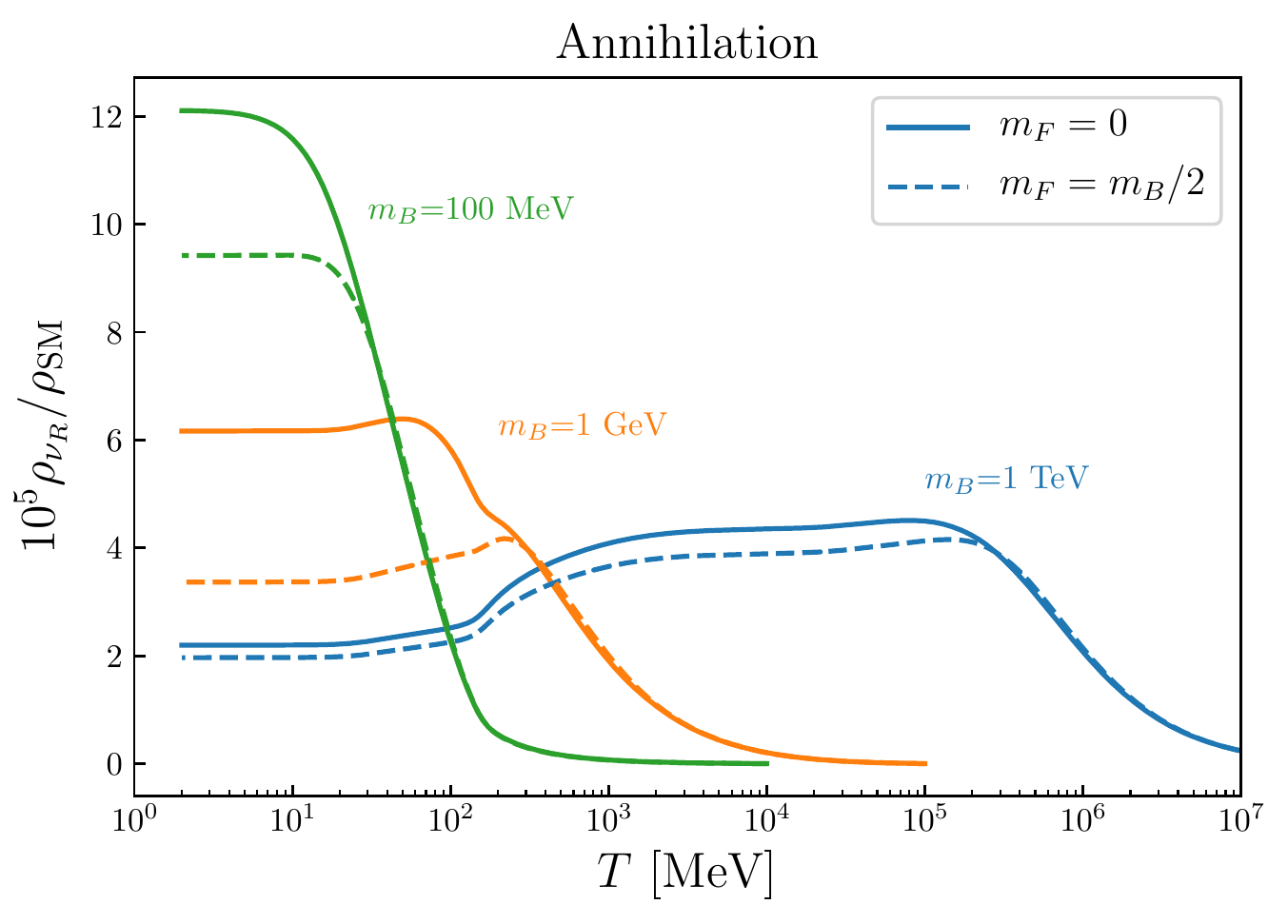}
    
    \includegraphics[width=0.48\textwidth]{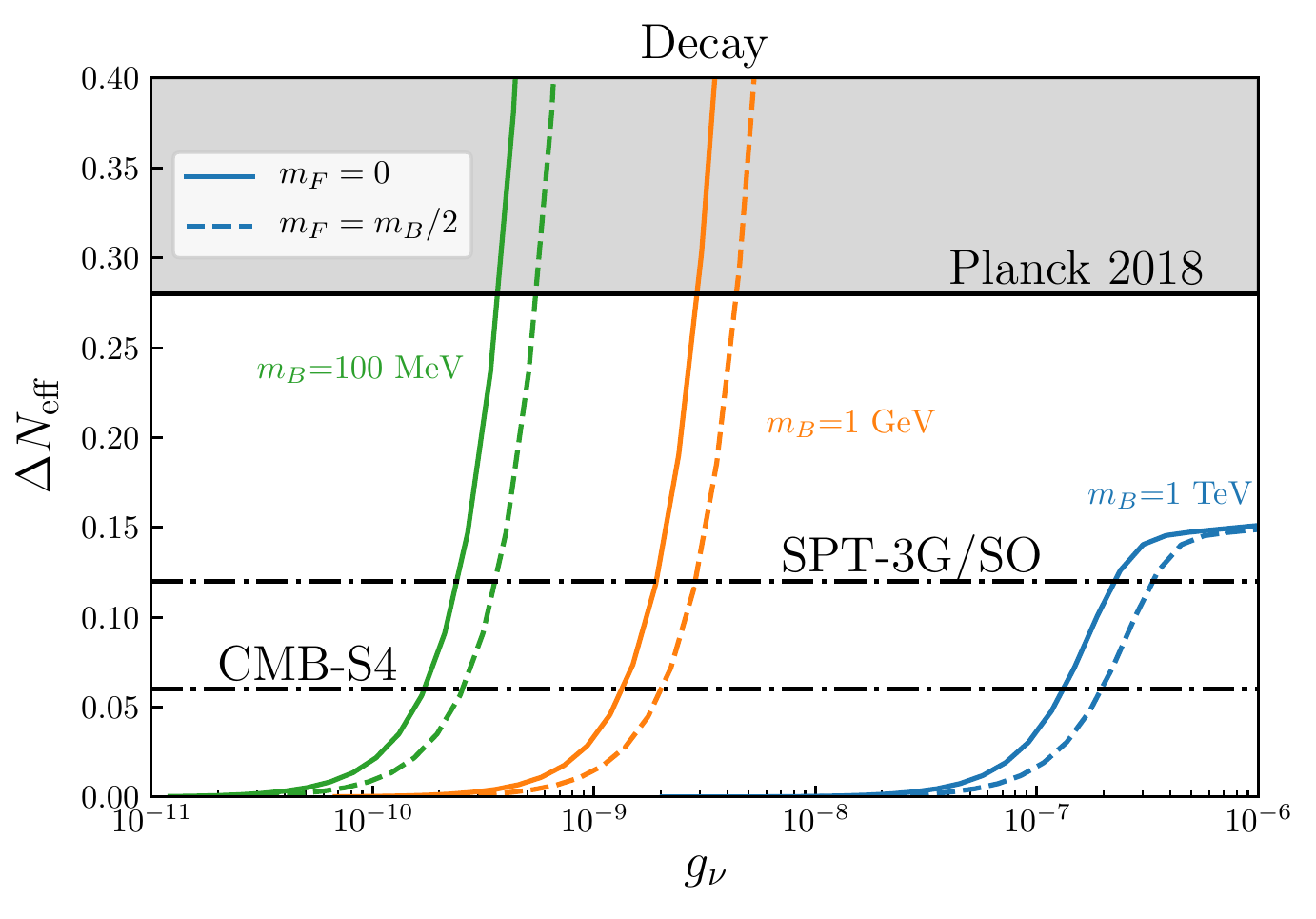}\includegraphics[width=0.48\textwidth]{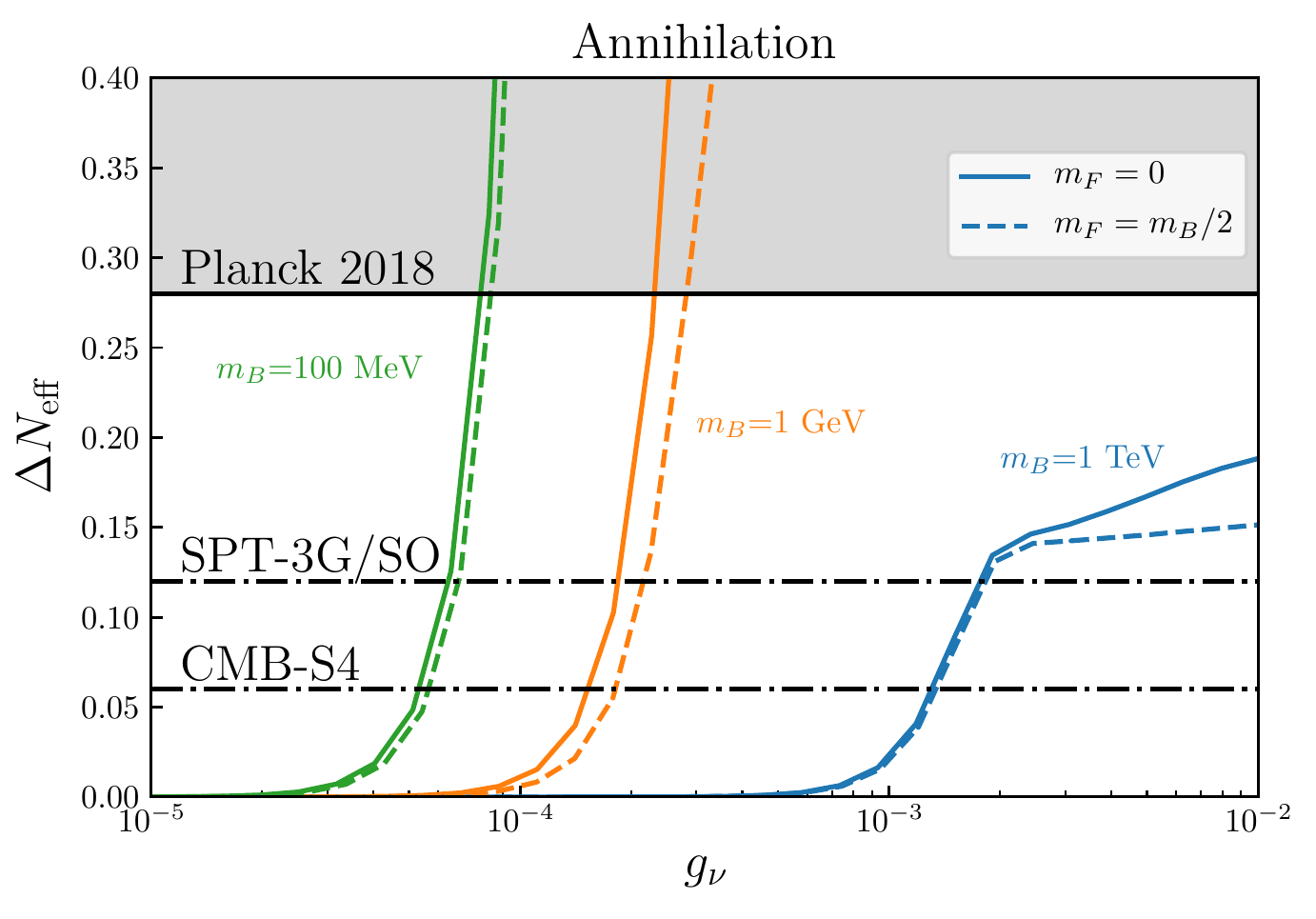}
    
    \caption{ Similar to Fig.~\ref{fig:N_eff_g_decay} but in order to illustrate the effect of  $\min(m_F,\, m_B)\neq 0$,  
    we compare curves of $m_F=m_B/2$ with  $m_F=0$, assuming subcases (I-1) and (III-2) in the left and right panels, respectively.
    Other relevant parameters are specified in the text. \label{fig:diff_m}}
\end{figure}

In the upper panels of Fig.~\ref{fig:N_eff_g_decay} we present the solutions
obtained from Eq.~(\ref{eq:m-91}) for several selected samples for
decay (left) and annihilation (right) processes. The former includes
four subcases: (I-1), (I-2), (II-1) and (III-1); and the later includes
two subcases: (II-2) and (III-2). Their collision terms are computed
according to Eq.~(\ref{eq:m-98}) with $C_{\nu_{R}}(T,\ 0)$ given
as follows:
\begin{eqnarray}
C_{\nu_{R}}^{\textrm{(I-1)}}(T,\ 0) & = & S|{\cal M}|^{2}\int d\Pi_{1}d\Pi_{3}d\Pi_{4}\frac{E_{3}}{e^{E_{1}/T}-1}\left[1-\frac{1}{e^{E_{4}/T}+1}\right](2\pi)^{4}\delta^{4},\label{eq:m-92}\\
C_{\nu_{R}}^{\textrm{(I-2)}}(T,\ 0) & = & S|{\cal M}|^{2}\int d\Pi_{1}d\Pi_{3}d\Pi_{4}\frac{E_{3}}{e^{E_{1}/T}+1}\left[1+\frac{1}{e^{E_{4}/T}-1}\right](2\pi)^{4}\delta^{4},\label{eq:m-93}\\
C_{\nu_{R}}^{\textrm{(II-1)}}(T,\ 0) & = & S|{\cal M}|^{2}\int d\Pi_{1}d\Pi_{3}d\Pi_{4}\frac{E_{3}}{e^{E_{1}/T}-1}(2\pi)^{4}\delta^{4},\label{eq:m-94}\\
C_{\nu_{R}}^{\textrm{(III-1)}}(T,\ 0) & = & S|{\cal M}|^{2}\int d\Pi_{1}d\Pi_{3}d\Pi_{4}\frac{E_{3}}{e^{E_{1}/T}+1}(2\pi)^{4}\delta^{4},\label{eq:m-95}\\
C_{\nu_{R}}^{\textrm{(II-2)}}(T,\ 0) & = & \int d\Pi_{1}d\Pi_{2}d\Pi_{3}d\Pi_{4}\frac{(2\pi)^{4}\delta^{4}E_{3}S|{\cal M}|^{2}}{\left(e^{E_{1}/T}-1\right)\left(e^{E_{2}/T}-1\right)},\label{eq:m-96}\\
C_{\nu_{R}}^{\textrm{(III-2)}}(T,\ 0) & = & \int d\Pi_{1}d\Pi_{2}d\Pi_{3}d\Pi_{4}\frac{(2\pi)^{4}\delta^{4}E_{3}S|{\cal M}|^{2}}{\left(e^{E_{1}/T}+1\right)\left(e^{E_{2}/T}-1\right)}.\label{eq:m-97}
\end{eqnarray}
Here $\delta^{4}$ is short for $\delta^{4}(p_{1}-p_{3}-p_{4})$
or $\delta^{4}(p_{1}+p_{2}-p_{3}-p_{4})$. For $S|{\cal M}|^{2}$,
we take the scalar results from Tab.~\ref{tab:Dominant-processes}.
Note that despite $S|{\cal M}|^2$ being the same for subcases (I-1) and (II-1), or for subcases (I-2) and (III-1), 
the above expressions of $C_{\nu_R}$ for these cases are different.
The initial particle mass,
$m_{1}$, should be either $m_{F}$ or $m_{B}$, as already specified
in Tab.~\ref{tab:Dominant-processes} for each subcase. 
We select in Fig.~\ref{fig:N_eff_g_decay} 
three representative values of $m_{1}$: 1 TeV, 1 GeV, and 100 MeV,
with  $g_{\nu}=10^{-8}$ ($2.8\times10^{-4})$, $1.6\times10^{-10}$
($4.4\times10^{-5}$), and $2\times10^{-11}$ ($1.8\times10^{-5}$)
in the left (right) panel, respectively. 
In Fig.~\ref{fig:N_eff_g_decay},  we set $m_{F}=0$ if $m_{F}<m_{B}$
or $m_{B}=0$ if $m_{B}<m_{F}$; the effect of nonzero $m_{F}$ or $m_{B}$ is shown in Fig.~\ref{fig:diff_m}.

In the lower panels of Fig.~\ref{fig:N_eff_g_decay}, we show the contribution to $N_{\rm eff}$ according to Eqs.~(\ref{eq:m-87})
or (\ref{eq:m-88}) as a function of  $g_{\nu}$ with
$m_{F}$ and $m_{B}$ being the same as in the upper panel. Results for subcases (II-1) and (III-1) are not
presented in the lower left panel because, as already suggested by
the upper left panel, they would be in between subcases (I-1)
and (I-2). 
We confront the results with current and future experimental bounds on $\Delta N_{\rm eff}$ from 
Planck 2018~\cite{Akrami:2018vks,Aghanim:2018eyx},
the Simons Observatory (SO)~\cite{Abitbol:2019nhf}, 
the South Pole Telescope (SPT-3G)~\cite{Benson:2014qhw},
and CMB-S4~\cite{Abazajian:2016yjj,Abazajian:2019eic}.
The Planck 2018 measurement gives $N_{{\rm eff}}=2.99\pm0.17$ (1$\sigma$) which after subtracting the $\nu_L$ 
contribution ($2.99-3.045=-0.055$) is recast as $\Delta N_{{\rm eff}}<0.17\times2-0.055=0.285$ at 2$\sigma$ C.L.
The SO and SPT-3G sensitivities are similar ($\Delta N_{{\rm eff}}<0.12$ at 2$\sigma$ C.L.),  labeled together as SO/SPT-3G. 
Finally, the future CMB-S4 limit is expected to reach 0.06, also at 
2$\sigma$ C.L.

\begin{figure}
    \centering
    
    \includegraphics[width=0.8\textwidth]{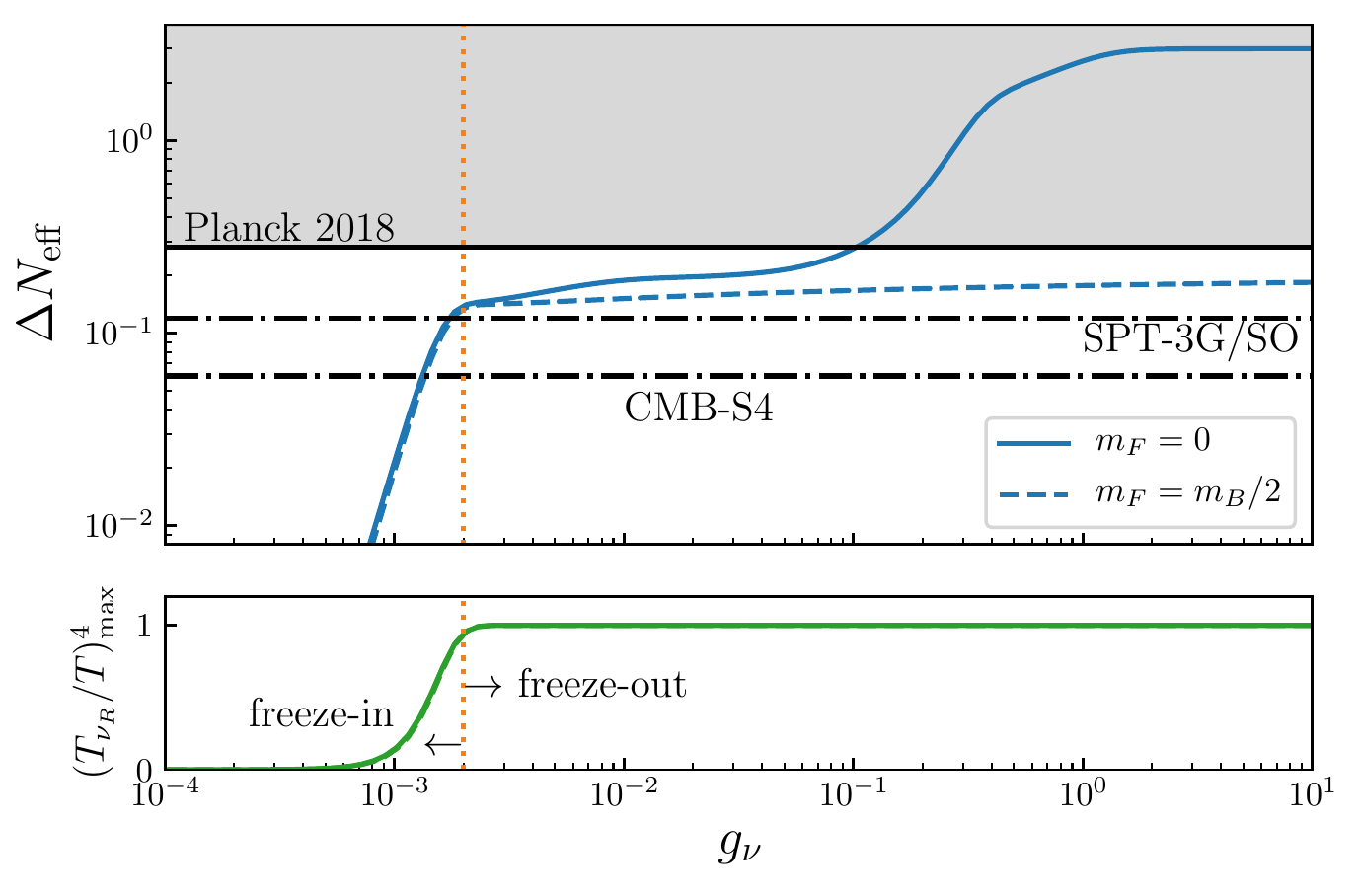}    
    \caption{Transition from the freeze-in to freeze-out regimes when $g_{\nu}$ increases to sufficiently large values. 
    The shown example takes $m_B=1$ TeV in subcase (III-2). In the lower panel, the maximal temperature ratio 
    $(T_{\nu_R}/T)^4_{\max}$ indicates whether $\nu_R$ had been in thermal equilibrium. 
    In the upper panel, the two plateaus at $\Delta N_{\rm eff}\approx 0.14$ and $\Delta N_{\rm eff}\approx 3$ correspond to $\nu_R$ decoupling above the electroweak scale and around the MeV scale,  respectively.
    The latter does not exist for the dashed curve because with $m_F=m_B/2$, the collision term becomes exponentially suppressed below the electroweak scale---see the text for more discussions. 
    \label{fig:freeze-in-out}}
\end{figure}

\begin{figure}[t]
\centering

\includegraphics[width=0.7\textwidth]{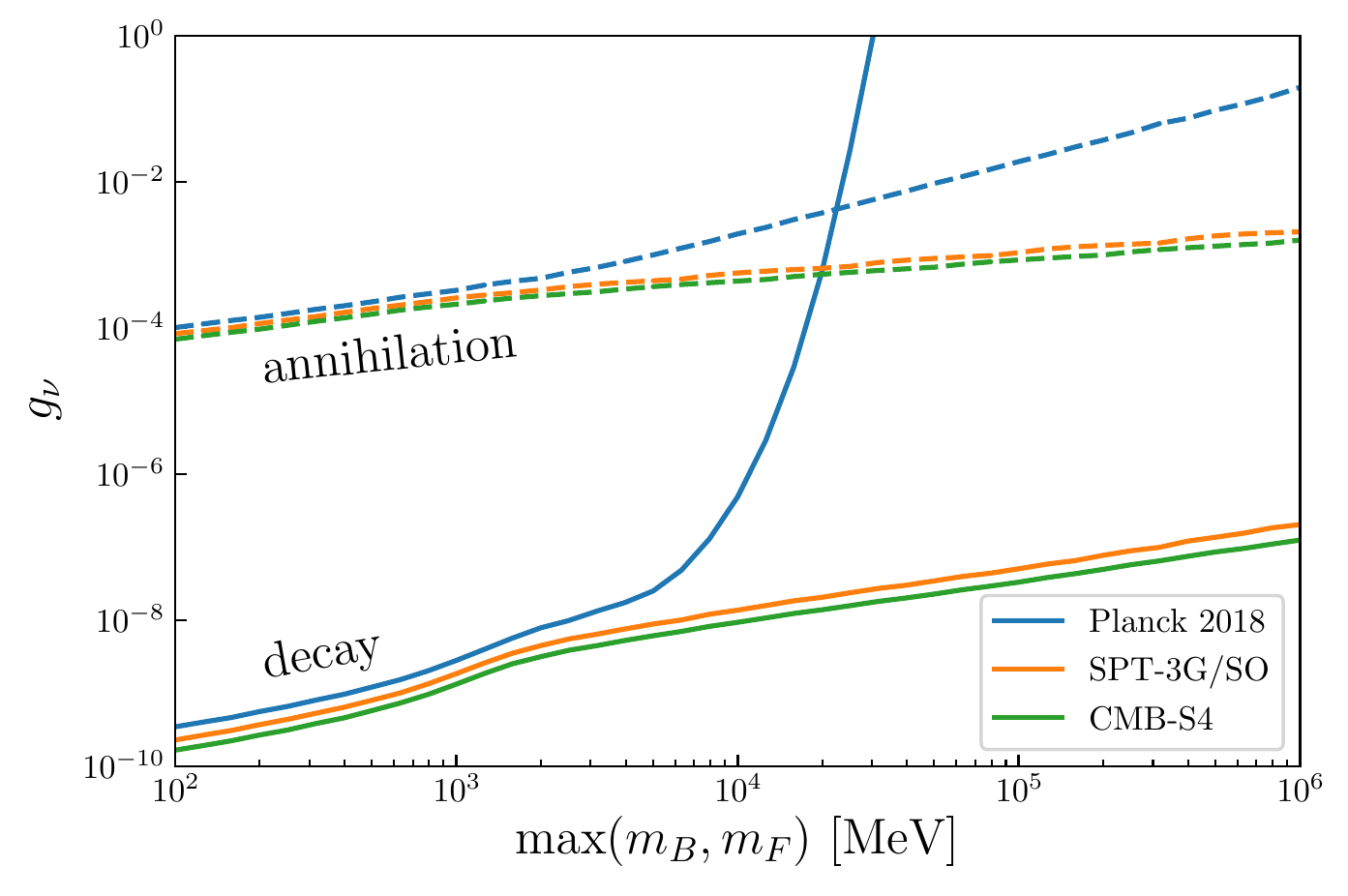}

\caption{Upper bounds on $g_{\nu}$ obtained from the requirement that $\Delta N_{{\rm eff}}$
does not exceed the current measurement of Planck 2018~\cite{Akrami:2018vks,Aghanim:2018eyx} or the sensitivity
of future CMB experiments including SO~\cite{Abitbol:2019nhf}, SPT-3G~\cite{Benson:2014qhw}, and CMB-S4~\cite{Abazajian:2016yjj,Abazajian:2019eic}.
\label{fig:g-m} 
}
\end{figure}

As shown in Fig.~\ref{fig:N_eff_g_decay}, for decay processes the production of $\nu_R$ is  most efficient 
when the temperature is lower than the initial particle mass $m_1$. 
Typically most $\nu_R$ are produced within $0.1m_1\lesssim T\lesssim m_1$. 
For annihilation processes, the production is  most efficient around $T\sim m_X$, the mass of the internal particle in the process.   
After that, the $\rho_{\nu_R}/\rho_{\rm SM}$ curves would remain stable  if the composition of the SM plasma was not changed.
However, at low temperatures due to  many heavy SM species annihilating or decaying into light ones, 
the comoving energy density of SM increases and hence $\rho_{\nu_R}/\rho_{\rm SM}$ decreases when $\nu_R$ 
is no longer effectively produced.
The most significant decrease in the curve appears during 100 MeV $\lesssim T\lesssim 1$ GeV, where $g_{\star}$ becomes substantially smaller. 
This feature holds for GeV or TeV masses, for lighter particles 
$\nu_R$ has not been produced yet in significant amounts.

The differences between dashed and solid curves in Fig.~\ref{fig:N_eff_g_decay} are caused 
by differences of $C_{\nu_R}$ in Eqs.~(\ref{eq:m-92})-(\ref{eq:m-97}), or more specifically, 
by the difference between Fermi-Dirac and Bose-Einstein statistics. The ``$\pm$'' and ``$\mp$'' signs 
in Eqs~(\ref{eq:m-21}) and (\ref{eq:m-81}) can lead to enhancement or suppression of $\rho_{\nu_R}$ by a factor of $R$ with 
$R\lesssim 1.5$ (decay) or  $R\lesssim 4$ (annihilation).
Consequently, the effect on $\Delta N_{\rm eff}$-$g_{\nu}$ in the lower panels is approximately a horizontal shift 
by a factor of $R^{1/2}$ (decay) or $R^{1/4}$ (annihilation) because in the freeze-in mechanism 
we have  $\Delta N_{\rm eff} \propto g_{\nu}^2$ and $\Delta N_{\rm eff} \propto g_{\nu}^4$ for  decay and 
annihilation  processes respectively. 
The effect of nonzero $\min(m_B,\, m_F)$ is quite similar, as shown in Fig.~\ref{fig:diff_m}. Taking subcases (I-1) and (III-2) as examples, in which $m_B$ is assumed to be larger than $m_F$, we plot curves for both $m_F=m_B/2$ and $m_F=0$. The difference can be accounted for also by the $R$ factor which is typically around 2 or 3, leading to a $R^{1/2}$ or $R^{1/4}$ horizontal shift  of the $\Delta N_{\rm eff}$-$g_{\nu}$ curves. 
Note that the case of $m_F=0$ could correspond to $F$ being the left-handed component of the Dirac neutrinos. 


Here we comment on a noteworthy behavior of large $g_\nu$ when $m_X$, the propagator mass in the annihilation case, is above the electroweak scale. For sufficiently large  $g_\nu$, $\nu_R$ can reach thermal equilibrium at a temperature well above the electroweak scale. If the initial particle mass $m_1=\min(m_B,\, m_F)$  is also above the electroweak scale, then at a lower (yet still above the electroweak scale) temperature $\nu_R$ will leave thermal equilibrium because the collision term is exponentially suppressed at $T\ll m_1$. Therefore, in this case, $\nu_R$ reaches and leaves thermal equilibrium at temperatures above the electroweak scale, leading to a constant $\Delta N_{\rm eff}\approx 0.14$~\cite{Abazajian:2019oqj,Luo:2020sho}.
If $m_1$ is below the electroweak scale, the decoupling temperature generally depends on $g_\nu$. As shown by the blue dashed and solid curve in the lower right panel in Fig.~\ref{fig:diff_m}, larger $g_\nu$ may or may not increase $\Delta N_{\rm eff}$, depending on whether $m_1$ is below or above the electroweak scale.

In Fig.~\ref{fig:freeze-in-out}, we further explore the dependence of $\Delta N_{\rm eff}$ on even larger $g_\nu$.
Here we take subcase (III-2) with $m_B=1$ TeV and $m_1=m_F=\{0, m_B/2\}$. For more general values of $m_1$ below  the electroweak scale, the result would be between the blue solid and dashed curves.
As has been expected, for larger $g_{\nu}$, the blue solid curve further increases and eventually reaches the maximal value ($\Delta N_{\rm eff}=3$) that $\nu_R$ could produce (in this case we assume $F$ is $\nu_L$); while the blue dashed curve is  insensitive to $g_\nu$, approximately keeping a constant value of $\Delta N_{\rm eff}$ at 0.14.

Fig.~\ref{fig:freeze-in-out} also shows explicitly the transition of the freeze-in to freeze-out regimes. 
Actually, for strong couplings $\nu_R$ had been in thermal equilibrium, thus its relic abundance depends on how late it would decouple from the SM plasma rather than how fast it was initially produced. As indicated by the lower panel, $(T_{\nu_R}/T)^4_{\max}$, defined as the maximal value of $(T_{\nu_R}/T)^4$ during the entire evolution, reaches $1$ when $g_{\nu}\gtrsim 2\times 10^{-3}$ (the orange dashed line). This is a good measure for the transition from the freeze-in to the freeze-out regime.





Finally, by requiring that the contribution of $\nu_R$ to $N_{\rm eff}$  does not exceed the current limit or future sensitivities of the aforementioned CMB experiments, we can obtain upper bounds on $g_\nu$. They are presented in Fig.~\ref{fig:g-m}, where we select subcases (I-1) and (III-2) for the decay and annihilation curves,  respectively. 
Here we set $\min(m_{B},\ m_{F})=0$ and $\max(m_{B},\ m_{F})\geq10^{2}$
MeV. The latter is to ensure that the calculation is not affected by $\nu_{L}$
decoupling. As previously discussed, for decay processes most $\nu_{R}$
are produced within $0.1\lesssim T/\max(m_{B},\ m_{F})\lesssim1$.
For $\min(m_{B},\ m_{F})=0$, we assume that $B$ and $F$ do not
contribute to $\Delta N_{{\rm eff}}$ significantly (e.g.\ $F$ may
be $\nu_{L}$).  One can also set $\min(m_{B},\ m_{F})$ to 10 MeV
for example to suppress their contributions to $\Delta N_{{\rm eff}}$.
This causes very insignificant changes in the final results.
As we have demonstrated in Figs.~\ref{fig:N_eff_g_decay} and \ref{fig:diff_m}, selecting other cases or using nonzero values of  $\min(m_B, m_F)$ typically increases or reduces $\Delta N_{\rm eff}$ by a factor of $R\approx 2\sim 4$ and hence the bounds on $g_{\nu}$ by a factor of $R^{1/2}$ or $R^{1/4}$. However, since the Planck 2018 limit on $\Delta N_{\rm eff}$ is above 0.14, for large masses the bounds can be weakened drastically and become more mass dependent. In fact, if $\nu_R$ production and decoupling (if it ever reached thermal equilibrium) are all well beyond the electroweak scale (this leads to $\Delta N_{\rm eff}\leq 0.14$),  Planck 2018 cannot provide a valid constraint on it. For the SO/SPT-3G and CMB-S4 curves, because these future experiments will be probing the freeze-in regime for large masses, the curves will not be significantly changed if  nonzero values of $\min(m_B, m_F)$ are used. Generally, we can draw the conclusion that for $\max(m_B, m_F)<1$ GeV, the current CMB measurement excludes $g_{\nu}\gtrsim 10^{-9}$  or $g_{\nu}\gtrsim 10^{-3}$ via the decay or  annihilation processes,  respectively. For larger masses, the Planck 2018 bounds are more mass-dependent (depending on both $\min(m_B, m_F)$  and $\max(m_B, m_F)$),  while  the SO/SPT-3G and CMB-S4 bounds mainly depend on $\max(m_B, m_F)$, where power-law extrapolations according to Eqs.~(\ref{eq:m-89})  and (\ref{eq:m-90}) can be used.

\section{Conclusion\label{sec:Conclusion}}
\noindent 
Dirac neutrinos with new interactions can have a measurable effect on the 
effective number of relativistic neutrino species $N_{\rm eff}$ in the early Universe, courtesy of a possible thermalization of the right-handed components $\nu_R$. We have computed here the effect of new vector and scalar interactions of right-handed neutrinos with new bosons and chiral fermions. Various special cases of this framework exist, depending on which particle is in equilibrium and which one is heavier, see Tab.~\ref{tab:Dominant-processes}. 
We focused on freeze-in of the right-handed neutrinos, and confronted the results with present and upcoming precise determinations of $\Delta N_{\rm eff}$. 

Approximate analytical results are given in Eqs.~(\ref{eq:m-89}) and (\ref{eq:m-90}); 
the outcome of a numerical solutions of the relevant equations is given in Figs.~\ref{fig:N_eff_g_decay} to \ref{fig:g-m}. 
For instance, if decay (scattering) of new particles is the dominating freeze-in process, limits on the new coupling constants of order 
$10^{-4}$ ($10^{-9}$) may be constrained for new particle masses around GeV. Chiral fermions being in equilibrium and massless can correspond to SM neutrinos. This also allows to consider the case of Dirac neutrino masses generated by the SM Higgs mechanism, 
which gives (see Fig.~\ref{fig:SM_Higgs}) 
$\Delta N_{\rm eff}^{\rm SM} \approx 7.5\times 10^{-12} \, (m_{\nu}/(0.1\, {\rm eV}))^{2}$. 

The results of this paper cover a wide range of possibilities, and demonstrate once more that cosmological measurements can constrain fundamental properties of particle physics, in particular neutrino physics. 

\begin{acknowledgments}
\noindent
We thank Laura Lopez-Honorez for helpful discussions. 
XJX is supported by the ``Probing dark matter with neutrinos'' ULB-ARC convention and by the F.R.S./FNRS under the Excellence of Science (EoS) project No. 30820817 - be.h ``The $H$ boson gateway to physics beyond the Standard Model''.
\end{acknowledgments}

\appendix

\section{Analytical results of 3-particle phase space integrals\label{sec:3particle}}
\noindent 
Since in many $1\rightarrow2$ processes the squared amplitudes $|{\cal M}|^{2}$
are energy-independent, it is useful to present the analytical results
of the following integrals:
\begin{eqnarray}
I^{(n)} & \equiv & \int d\Pi_{1}d\Pi_{3}d\Pi_{4}(2\pi)^{4}\delta^{4}(p_{1}-p_{3}-p_{4})e^{-E_{1}/T},\label{eq:m-72}\\
I^{(\rho)} & \equiv & \int d\Pi_{1}d\Pi_{3}d\Pi_{4}(2\pi)^{4}\delta^{4}(p_{1}-p_{3}-p_{4})e^{-E_{1}/T}E_{3},\label{eq:m-73}
\end{eqnarray}
where $m_{1}\neq0$ and $m_{3}=m_{4}=0$. 

The results are
\begin{eqnarray}
I^{(n)} & \equiv & \frac{1}{32\pi^{3}}m_{1}TK_{1}\left(\frac{m_{1}}{T}\right),\label{eq:m-74}\\
I^{(\rho)} & \equiv & \frac{1}{64\pi^{3}}m_{1}^{2}TK_{2}\left(\frac{m_{1}}{T}\right),\label{eq:m-75}
\end{eqnarray}
where $K_{1}$ and $K_{2}$ are $K$-type Bessel functions of order
$1$ and $2$ respectively. 
Next we derive these two analytical results.

First, we substitute $(2\pi)^{3}\delta^{3}(\boldsymbol{p}_{1}-\boldsymbol{p}_{3}-\boldsymbol{p}_{4})=\int e^{i(\boldsymbol{p}_{1}-\boldsymbol{p}_{3}-\boldsymbol{p}_{4})\cdot\boldsymbol{\lambda}}d^{3}\boldsymbol{\lambda}$
and $d\Pi_{i}=\frac{p_{i}^{2}dp_{i}d\Omega_{i}}{(2\pi)^{3}2E_{i}}$
in Eqs.~(\ref{eq:m-72}) and (\ref{eq:m-73}): 
\begin{equation}
I\equiv\frac{1}{(2\pi)^{9}}\int\frac{p_{1}^{2}dp_{1}}{2E_{1}}\frac{p_{3}^{2}dp_{3}}{2E_{3}}\frac{p_{4}^{2}dp_{4}}{2E_{4}}2\pi\delta(E_{1}-p_{3}-p_{4})e^{-E_{1}/T}UI_{\Omega},\label{eq:m-100}
\end{equation}
where $U=E_{3}$ for $I^{(\rho)}$ or $1$ for $I^{(n)}$, and $I_{\Omega}$
contains the angular part of the integral:
\begin{eqnarray}
I_{\Omega} & = & \int d^{3}\boldsymbol{\lambda}\int d\Omega_{1}e^{i\boldsymbol{p}_{1}\cdot\boldsymbol{\lambda}}\int d\Omega_{3}e^{-i\boldsymbol{p}_{3}\cdot\boldsymbol{\lambda}}\int d\Omega_{4}e^{-i\boldsymbol{p}_{4}\cdot\boldsymbol{\lambda}}.\label{eq:m-101}
\end{eqnarray}
Since $\int d\Omega_{i}e^{\pm i\boldsymbol{p}_{i}\cdot\boldsymbol{\lambda}}=\int dc_{i}d\phi_{i}e^{\pm ip_{i}\lambda c_{i}}=4\pi\frac{\sin(p_{i}\lambda)}{p_{i}\lambda}$,
we further get 
\begin{eqnarray}
I_{\Omega} & = & (4\pi)^{3}\int d^{3}\boldsymbol{\lambda}\frac{\sin(p_{1}\lambda)}{p_{1}\lambda}\frac{\sin(p_{3}\lambda)}{p_{3}\lambda}\frac{\sin(p_{4}\lambda)}{p_{4}\lambda}\nonumber \\
 & = & (4\pi)^{4}\int_{0}^{\infty}\frac{d\lambda}{p_{1}p_{3}p_{4}\lambda}\sum_{\eta_{1},\eta_{3},\eta_{4}}\frac{-\eta_{1}\eta_{3}\eta_{4}}{8}\sin\left(\eta_{1}p_{1}\lambda+\eta_{3}p_{3}\lambda+\eta_{4}p_{4}\lambda\right)\nonumber \\
 & = & \frac{32\pi^{5}}{p_{1}p_{3}p_{4}}\left[\frac{p_{1}-p_{3}+p_{4}}{|p_{1}-p_{3}+p_{4}|}+\frac{p_{1}+p_{3}-p_{4}}{|p_{1}+p_{3}-p_{4}|}-\frac{p_{1}-p_{3}-p_{4}}{|p_{1}-p_{3}-p_{4}|}-\frac{p_{1}+p_{3}+p_{4}}{|p_{1}+p_{3}+p_{4}|}\right],\label{eq:m-102}
\end{eqnarray}
where in the second line $\eta_{i}=\pm1$ denotes positive/negative
signs, and in the last line we have used $\int\frac{d\lambda}{\lambda p}\sin(\lambda p)=1/|p|$.

Using Eq.~(\ref{eq:m-102}), it is straightforward to integrate out
$p_{3}$ and $p_{4}$ in Eq.~(\ref{eq:m-100}), leading to
\begin{eqnarray}
I^{(n)} & = & \frac{1}{(2\pi)^{9}}\int\frac{16\pi^{6}p_{1}^{2}}{E_{1}}e^{-E_{1}/T}dp_{1}=\frac{1}{32\pi^{3}}\int_{m_{1}}^{\infty}p_{1}e^{-E_{1}/T}dE_{1},\label{eq:m-103}\\
I^{(\rho)} & = & \frac{1}{(2\pi)^{9}}\int8\pi^{6}p_{1}^{2}e^{-E_{1}/T}dp_{1}=\frac{1}{64\pi^{3}}\int_{m_{1}}^{\infty}p_{1}E_{1}e^{-E_{1}/T}dE_{1}.\label{eq:m-104}
\end{eqnarray}
The above integrals can be expressed in terms of the Bessel functions, as already given in Eqs.~(\ref{eq:m-74}) and (\ref{eq:m-75}).

\section{ Monte-Carlo integration of general  collision terms\label{sec:Numerical-collision}}
\noindent 
In this appendix, we introduce the techniques we use to numerically
evaluate the phase space integrals of collision terms. The method
is based on Monte-Carlo integration and in principle applies to any
$m\rightarrow n$ ($m$, $n=$1, 2, 3,$\cdots$) processes. 

Consider the following integral
\begin{equation}
I[{\cal F}]\equiv\int d\Pi_{1}d\Pi_{2}\cdots d\Pi_{m+n}(2\pi)^{4}\delta^{4}(p_{1}+p_{2}+\cdots p_{m}-p_{m+1}-\cdots p_{m+n}){\cal F}(p_{1},\ p_{2},\ \cdots),\label{eq:m}
\end{equation}
where $p_{1}$, $p_{2}$, $\cdots$, $p_{m}$ ($p_{m+1}$, $\cdots$,
$p_{m+n}$) are momenta of initial (final) particles, 
\begin{equation}
d\Pi_{i}=\frac{d^{3}p_{i}}{(2\pi)^{3}2E_{i}},\label{eq:m-1}
\end{equation}
and ${\cal F}$ is a general function of all the momenta. For simplicity,
we  denote $p_{1}+p_{2}+\cdots p_{m}-p_{m+1}-\cdots-p_{m+n-2}$ by
$q$, and the last two momenta $p_{m+n}$ and $p_{m+n-1}$ by $p_{\bar{1}}$
and $p_{\bar{2}}$, respectively. 

There are two technical problems in the Monte-Carlo integration that
we need to deal with properly, otherwise the Monte-Carlo integration
would converge very slowly. The first one concerns the $\delta$ function,
which will be removed by integrating out some part of the momenta.
The second problem is that the integration domain is infinitely large,
which can be avoided by a proper transformation of variables.

To remove the $\delta$ function, we first integrate out $\boldsymbol{p}_{\bar{1}}$
so that 
\begin{equation}
I=\int d\Pi_{1}d\Pi_{2}\cdots d\Pi_{m+n-1}\frac{2\pi}{2E_{\bar{1}}}\delta(E_{q}-E_{\bar{2}}-E_{\bar{1}}){\cal F},\label{eq:m-2}
\end{equation}
where $E_{q}$, $E_{\bar{2}}$ and $E_{\bar{1}}$ are the energies
of the on-shell momenta $q$, $p_{\bar{1}}$ and $p_{\bar{2}}$, respectively.
Note that since $\boldsymbol{p}_{\bar{1}}$ has already been integrated
out in Eq.~(\ref{eq:m-2}), instead of being a function of $\boldsymbol{p}_{\bar{1}}$,
$E_{\bar{1}}$ should be interpreted as a function of $\boldsymbol{q}$
and $\boldsymbol{p}_{\bar{2}}$:
\begin{equation}
E_{\bar{1}}=\sqrt{m_{\bar{1}}^{2}+|\boldsymbol{q}-\boldsymbol{p}_{\bar{2}}|^{2}}.\label{eq:m-3}
\end{equation}
Next, we integrate out $|\boldsymbol{p}_{\bar{2}}|$ in Eq.~(\ref{eq:m-2})
and obtain
\begin{equation}
I=\int d\Pi_{1}d\Pi_{2}\cdots d\Pi_{m+n-2}\frac{|\boldsymbol{p}_{\bar{2}}|^{2}dc_{\bar{2}}d\phi_{\bar{2}}}{(2\pi)^{3}2E_{\bar{2}}}\frac{2\pi}{2E_{\bar{1}}}J^{-1}{\cal F}\Theta,\label{eq:m-2-1}
\end{equation}
where $c_{\bar{2}}=\cos\theta_{\bar{2}}$, $\theta_{\bar{2}}$ and
$\phi_{\bar{2}}$ are the polar and azimuthal angles in a spherical
coordinate system with the zenith direction aligned with $\boldsymbol{q}$
(hence $\boldsymbol{p}_{\bar{2}}\cdot\boldsymbol{q}=|\boldsymbol{p}_{\bar{2}}||\boldsymbol{q}|c_{\bar{2}}$),
and  
\begin{equation}
J^{-1}=\left|\frac{\partial(E_{\bar{2}}+E_{\bar{1}})}{\partial|\boldsymbol{p}_{\bar{2}}|}\right|^{-1}=\left|\frac{|\boldsymbol{p}_{\bar{2}}|}{E_{\bar{2}}}+\frac{|\boldsymbol{p}_{\bar{2}}|-|\boldsymbol{q}|c_{\bar{2}}}{E_{\bar{1}}}\right|^{-1},\label{eq:m-12}
\end{equation}
according to the property of $\delta$ function: $\delta(g(x))=\delta(x-x_{0})\left|g'(x_{0})\right|^{-1}$
with $x_{0}$ being a root of  $g(x_{0})=0$. 

The Heaviside theta function  $\Theta$ takes either 1 or 0 depending
on whether $q^{\mu}$ and $c_{\bar{2}}$ lead to  physical kinematics
or not. Technically, it is computed as follows:
\begin{equation}
\Theta=\begin{cases}
1 & {\rm if}\ q^{2}>(m_{\bar{1}}+m_{\bar{2}})^{2}\ \&\ \Delta>0\\
0 & {\rm otherwise}
\end{cases}\,,\label{eq:m-79}
\end{equation}
where
\begin{equation}
\Delta\equiv m_{\bar{2}}^{4}+\left(m_{\bar{1}}^{2}-q^{2}\right){}^{2}-2m_{\bar{2}}^{2}\left[q^{2}+2\left(1-c_{\bar{2}}^{2}\right)|\boldsymbol{q}|^{2}+m_{\bar{1}}^{2}\right].\label{eq:m-78}
\end{equation}
Note that in the above expression $q^{2}=E_{q}^{2}-|\boldsymbol{q}|^{2}$
is different from $|\boldsymbol{q}|^{2}$. The condition $q^{2}>(m_{\bar{1}}+m_{\bar{2}})^{2}$
enforces that $q$ provides sufficient energy to generate particles
$\bar{2}$ and $\bar{1}$. This can be derived in the center-of-mass
frame of particles $\bar{2}$ and $\bar{1}$, where $\boldsymbol{q}=0$
and it is obvious that near the threshold both particles should be
almost at rest. Slightly above the threshold, we need $E_{q}$ to
be slightly larger than $m_{\bar{1}}+m_{\bar{2}}$ to produce the
two particles. So in the center-of-mass frame, $E_{q}>m_{\bar{1}}+m_{\bar{2}}$
is necessary and sufficient for $q$ to produce the two particles.
In other frames with nonzero values of $|\boldsymbol{q}|$, by applying
a Lorentz transformation, we get $q^{2}-(m_{\bar{1}}+m_{\bar{2}})^{2}>0$.
The other requirement $\Delta>0$ puts a further constraint on the
angles, which will be derived in Eq.~(\ref{eq:m-5}).

Next, we need to reconstruct $\boldsymbol{p}_{\bar{2}}$ from given
values of $E_{q}$, $\boldsymbol{q}$, and $\theta_{\bar{2}}$. In
principle, $|\boldsymbol{p}_{\bar{2}}|$ in Eq.~(\ref{eq:m-2-1})
should be interpreted as an implicit function of these quantities
and $\phi_{\overline{2}}$. However, $\phi_{\overline{2}}$ turns out to
be irrelevant here. 

Given $E_{q}$, $\boldsymbol{q}$, and $c_{\bar{2}}$, $|\boldsymbol{p}_{\bar{2}}|$
is determined by
\begin{equation}
E_{q}=\sqrt{m_{\bar{2}}^{2}+|\boldsymbol{p}_{\bar{2}}|^{2}}+\sqrt{m_{\bar{1}}^{2}+|\boldsymbol{q}-\boldsymbol{p}_{\bar{2}}|^{2}},\label{eq:m-4}
\end{equation}
which can be solved  as a quadratic equation of  $|\boldsymbol{p}_{\bar{2}}|$
and gives 
\begin{equation}
|\boldsymbol{p}_{\bar{2}}|=\frac{c_{\bar{2}}|\boldsymbol{q}|\left(q^{2}-m_{\bar{1}}^{2}+m_{\bar{2}}^{2}\right)+E_{q}\sqrt{\Delta}}{2\left(E_{q}^{2}-c_{\bar{2}}^{2}|\boldsymbol{q}|^{2}\right)}.\label{eq:m-5}
\end{equation}
Eq.~(\ref{eq:m-5}) implies that $\Delta$ cannot be negative otherwise
Eq.~(\ref{eq:m-4}) would have no real solution. This sets a constraint
on $c_{\bar{2}}$. As can be seen from Eq.~(\ref{eq:m-78}), for a fixed
value of $q^{2}$, one can boost $|\boldsymbol{q}|^{2}$ to an arbitrarily
large value so that the $\left(1-c_{\bar{2}}^{2}\right)|\boldsymbol{q}|^{2}$
term is dominant and leads to $\Delta<0$, unless $1-c_{\bar{2}}^{2}$
is suppressed. So generally speaking, for very large $|\boldsymbol{q}|^{2}$
and nonzero $m_{\bar{2}}^{2}$, the physically allowed region for
$1-c_{\bar{2}}^{2}$ is small. This feature could be used to improve
the the efficiency of Monte-Carlo integration by limiting the sampling
space of $c_{\bar{2}}^{2}$, though it has not been implemented
in our code. 

Once $|\boldsymbol{p}_{\bar{2}}|$ is determined from Eq.~(\ref{eq:m-5}),
we can readily compute $E_{\bar{2}}$, $E_{\bar{1}}$, and $|\boldsymbol{p}_{\bar{1}}|$. 

The second problem concerns the infinitely large domain of integration
(each $|\boldsymbol{p}_{i}|$ is integrated from 0 to $\infty$).
We make the following variable transformation for each $|\boldsymbol{p}_{i}|$:
\begin{equation}
x_{i}\equiv\exp\left(-|\boldsymbol{p}_{i}|/\Lambda_{i}\right),\ \ {\rm or}\ \ |\boldsymbol{p}_{i}|=-\Lambda_{i}\log(x_{i}),\label{eq:m-86}
\end{equation}
 and integrate $x_{i}$ from 0 to 1. In our code we usually take $\Lambda_{i}=4T_{i}$,
which usually leads to efficient convergence of the Monte-Carlo integration.
The transformation also generates another Jacobian:
\begin{equation}
    J_i\equiv dx_i/dp_i=-x_i/\Lambda\,,
    \label{eq:m-Jx}
\end{equation}
 which should be included in the integration via $dp_i\rightarrow dx_i/J_i$.  \\

In summary, the Monte-Carlo integration of $I$ can be implemented
as follows:
\begin{itemize}
\item Randomly generate values of $(x_i,\ c_i,\ \phi_i)$ with $i=1,\cdots , n+m-2$, $x_i \in (0,\ 1)$, $c_i \in (-1,\ 1)$, and   $\phi_i \in (0,\ 2\pi)$;
\item Construct the spatial parts of the first $n+m-2$ momenta ($\boldsymbol{p}_{1}$,
$\boldsymbol{p}_{2}$, $\cdots$, $\boldsymbol{p}_{n+m-2}$) from $(x_i,\ c_i,\ \phi_i)$;
\item Compute their respective energies $E_{1},E_{2},\cdots,E_{n+m-2}$
according to the on-shell condition;
\item Construct $q=(E_{q},\ \boldsymbol{q})$ with $E_{q}=\sum_{i=1}^{n+m-2}E_{i}$
and $\boldsymbol{q}=\sum_{i=1}^{n+m-2}\boldsymbol{p}_{i}$;
\item Randomly generate $c_{\bar{2}}$ and $\phi_{\bar{2}}$ in Eq.~(\ref{eq:m-2-1});
\item Compute $|\boldsymbol{p}_{\bar{2}}|$ according to Eq.~(\ref{eq:m-5})
so that the second last momentum $p_{\bar{2}}=(E_{\bar{2}},\ \boldsymbol{p}_{\bar{2}})$
can be reconstructed;
\item Reconstruct the last momentum according to $p_{\bar{1}}=q-p_{\bar{2}}$;
\item Evaluate the integrand in Eq.~(\ref{eq:m-2-1}) and proceed with the
standard Monte-Carlo procedure\footnote{In principle, one can also apply more advanced methods such as adaptive
Monte Carlo integration, but we find such methods in our case often
lead to biased results when the number of samples is not sufficiently
large.}. Note that  in addition to the Jacobian in Eq.~(\ref{eq:m-12}), there is also another Jacobian $J_i$ in Eq.~(\ref{eq:m-Jx}) that needs to be included.
\end{itemize}

\subsection{Example: $1\rightarrow2$ processes}
\noindent
As the simplest example, let us apply the above method to $1\rightarrow 2$ processes, 
\begin{equation}
    I\equiv\int d\Pi_{1}d\Pi_{2}d\Pi_{3}(2\pi)^{4}\delta^{4}(p_{1}-p_{2}-p_{3}){\cal F}(p_{1},\ p_{2},\ p_{3})\,.\label{eq:m-I3}
\end{equation}
Following the above notation, the $q$ momentum is identical to $p_{1}$ and hence $q^{2}=m_{1}^{2}$ which implies that in the $\Theta$ function the $q^{2}>(m_{2}+m_{3})^{2}$ condition (equivalent to $m_1>m_2+m_3$) can be ignored. The integral is computed as follows:
\begin{equation}
    I=\left\langle \frac{|\boldsymbol{p}_{1}|^{2}}{(2\pi)^{3}2E_{1}}\frac{|\boldsymbol{p}_{2}|^{2}}{(2\pi)^{3}2E_{2}}\frac{2\pi}{2E_{3}}J^{-1}J_{1}{\cal F}\Theta\right\rangle V,\label{eq:m-I3-1}
\end{equation}
where $\langle\rangle$ stands for the mean value after a large number
of evaluations of the inside quantity, $J_{1}$ is given in Eq.~(\ref{eq:m-Jx}), and
$V=1\times2^{2}\times(2\pi)^{2}$ is the volume of the sampling space:
$x_{1}\in(0,\ 1)$, $c_{1,2}\in(-1,\ 1)$, $\phi_{1,2}\in(0,\ 2\pi)$. 
Let us apply the the Monte-Carlo method to Eq.~(\ref{eq:m-73}), which has a known analytical result.
Taking $T=m_1=1$ GeV and assuming other particles are massless, 
the Bessel-form expression in Eq.~(\ref{eq:m-73}) gives $I=8.188\times10^{-4}\ {\rm GeV}^3$. 
Performing the Monte-Carlo evaluation of Eq.~(\ref{eq:m-I3-1}) with $10^7$ samples for ten times,  
we get $I/(10^{-4}\ {\rm GeV}^3)=$ \{8.196,  8.197,  8.203,  8.190,  8.164,  8.174,  8.186,  8.176,
8.195,  8.185\}, which is consistent with the analytical result.
Each evaluation with $10^7$ samples takes about three seconds using our code currently implemented in {\tt Python}.

\subsection{Example: $2\rightarrow2$ processes}
\noindent 
Consider a $2\rightarrow2$ process with the kinematics $p_{1}+p_{2}=p_{3}+p_{4}$
and $q=p_{1}+p_{2}$. In this case, we have 
\begin{equation}
I=\int d\Pi_{1}d\Pi_{2}\frac{|\boldsymbol{p}_{3}|^{2}dc_{3}d\phi_{3}}{(2\pi)^{3}2E_{3}}\frac{2\pi}{2E_{4}}J^{-1}{\cal F}\Theta,\label{eq:m-13}
\end{equation}
where ${\cal F}$ contains statistical distribution functions and
a scattering amplitude. The scattering amplitude usually can be expressed
in terms of $p_{1}\cdot p_{2}$, $p_{1}\cdot p_{3}$ and $p_{2}\cdot p_{3}$.
If it contains scalar products of $p_{4}$, then we can replace $p_{4}$
with $p_{1}+p_{2}-p_{3}$. For example, $p_{1}\cdot p_{4}$ can be
written as $p_{1}\cdot(p_{1}+p_{2}-p_{3})=m_{1}^{2}+p_{1}\cdot p_{2}-p_{1}\cdot p_{3}$. 

To facilitate the calculation of scalar products, it would be better
to define all the polar angles (i.e.\ $\theta$'s) with respective
to $\boldsymbol{q}$. But since $\boldsymbol{q}$ is constructed from
$\boldsymbol{p}_{1}$ and $\boldsymbol{p}_{2}$, such definitions
would be conceptually confusing. We perform the variable transformation:
$(\boldsymbol{p}_{1},\ \boldsymbol{p}_{2})\rightarrow(\boldsymbol{q},\ \boldsymbol{p}_{2})=(\boldsymbol{p}_{1}+\boldsymbol{p}_{2},\ \boldsymbol{p}_{2})$
to avoid this confusion. Since the Jacobian of this transformation
is $1$, after the transformation Eq.~(\ref{eq:m-13}) becomes 
\begin{equation}
I=\int\frac{|\boldsymbol{q}|^{2}d|\boldsymbol{q}|dc_{q}d\phi_{q}}{(2\pi)^{3}2E_{q}}\frac{|\boldsymbol{p}_{2}|^{2}d|\boldsymbol{p}_{2}|dc_{2}d\phi_{2}}{(2\pi)^{3}2E_{2}}\frac{|\boldsymbol{p}_{3}|^{2}dc_{3}d\phi_{3}}{(2\pi)^{3}2E_{3}}\frac{2\pi}{2E_{4}}J^{-1}{\cal F}\Theta,\label{eq:m-14}
\end{equation}
where $c_{2}=\cos\theta_{2}$ and $\theta_{2}$ is defined as the
angle between $\boldsymbol{p}_{2}$ and $\boldsymbol{q}$.

With the proper definition of $c_{2}$ (similar to $c_{3}$), we have
\begin{equation}
q\cdot p_{2}=E_{q}E_{2}-|\boldsymbol{q}||\boldsymbol{p}_{2}|c_{2},\ \ q\cdot p_{3}=E_{q}E_{3}-|\boldsymbol{q}||\boldsymbol{p}_{3}|c_{3},\label{eq:m-15}
\end{equation}
and 
\begin{equation}
p_{2}\cdot p_{3}=E_{2}E_{3}-|\boldsymbol{p}_{2}||\boldsymbol{p}_{3}|\left[s_{2}s_{3}\cos(\phi_{2}-\phi_{3})+c_{2}c_{3}\right],\label{eq:m-16}
\end{equation}
where $(s_{2},\ s_{3})\equiv(\sin\theta_{2},\ \sin\theta_{3})$. 
From Eqs.~(\ref{eq:m-15}) and (\ref{eq:m-16}), it is straightforward
to obtain any scalar products of $p_{1}$, $p_{2}$, $p_{3}$, and
$p_{4}$. 

It is also known that in the MB approximation, the collision terms
of contact interactions of four massless fermions are analytically
calculable. For example, given
\begin{equation}
{\cal F}=\exp(-E_{1}/T)\exp(-E_{2}/T)(p_{1}\cdot p_{2})(p_{3}\cdot p_{4}),\label{eq:m-17}
\end{equation}
the analytical result is (see Tab.~III in Ref.~\cite{Luo:2020sho}):
\begin{equation}
I=\frac{3T^{8}}{8\pi^{5}}\approx1.225\times10^{-3}\ T^{8}.\label{eq:m-18}
\end{equation}
Performing the Monte-Carlo integration described above with $10^{6}$
samples, we find that the numerical factor typically varies from $1.22\times10^{-3}$
to $1.23\times10^{-3}$, which is in agreement with the analytical
result.

\bibliographystyle{JHEP}
\bibliography{ref}

\end{document}